\documentclass[aps,prd,letterpaper]{revtex4}

\usepackage{latexsym}
\usepackage{epsfig}
\usepackage{scrextend}
 \usepackage[french]{babel}

\usepackage{url}
\usepackage{color}

\definecolor{darkblue}{rgb}{0,0,0.63} 
\usepackage[colorlinks,linkcolor=darkblue,citecolor=darkblue,urlcolor=darkblue]{hyperref}
\usepackage{latexsym}
\usepackage{epsfig}
\usepackage{scrextend}
\usepackage{lipsum}   
\usepackage{graphicx}
\usepackage{pdfpages}
\usepackage[title]{appendix}

\usepackage{natbib}

\usepackage{tikz}
\usetikzlibrary{positioning,arrows,decorations.pathmorphing,arrows.meta}

\usepackage{xcolor}
\usepackage{amssymb}
\usepackage{amsmath}
\usepackage{graphicx}

 \definecolor{blue}{RGB}{7,80,201}
 \definecolor{red}{RGB}{200,20,1}

\def\be{\begin{equation}}
\def\ee{\end{equation}}
\def\bea{\begin{eqnarray}}
\def\eea{\end{eqnarray}}
\def\ba{\begin{array}} 
\def\ea{\end{array}}
\def\bc{\begin{center}}
\def\ec{\end{center}}

\def\ghost#1{}

\def\simge{\mathrel{%
   \rlap{\raise 0.511ex \hbox{$>$}}{\lower 0.511ex \hbox{$\sim$}}}}
\def\simle{\mathrel{
   \rlap{\raise 0.511ex \hbox{$<$}}{\lower 0.511ex \hbox{$\sim$}}}}

\def\dis{\displaystyle}

\begin{document}

\title{\boldmath \Large  

$ \dis
\ba{c}
\hbox{Hyperbolic form factors for Yukawa interactions}, 
\vspace{3mm} \\ 
\hbox{and applications to the Earth} 
\ea 
$
\vspace{10mm}\\
}

\author{{\sc P}{ierre} {\sc Fayet}
\vspace{4mm} \\ \small }

\affiliation{Laboratoire de physique de l'\'Ecole normale sup\'erieure\vspace{.5mm}\\ 
ENS-PSL, CNRS, Sorbonne Univ., Univ.\,\,Paris Cit\'e, Paris, France
\,\footnote{\ pierre.fayet@phys.ens.fr}
\vspace{.5mm}\\
\hbox{and \,Centre de physique th\'eorique, \'Ecole polytechnique, IPP, Palaiseau, France}
\vspace{6mm}}

\date{March 22, 2026}

\begin{abstract}

\vspace{10mm}

We define  the hyperbolic form factor of a density distribution 
\vspace{.1mm}
as its bilateral Laplace transform, 
related by duality or analytic continuation to its ordinary form factor.
\vspace{-.7mm}
\,For a sphere it is 
given by $\Phi(x \!= \! kR) =\langle \,\cosh \,\vec k.\vec r\,\rangle=  \langle \,\frac{\sinh kr}{kr}\,\rangle$, expanded as $\,\sum_0^\infty \frac{x^{2n}}{(2n+1)! }\, \frac{\langle r^{2n}\rangle}{R^{2n}}$,
\vspace{-.9mm}
and similarly for the form factor $ \langle \,\frac{\sin kr}{kr}\,\rangle$. It is also obtained from the bilateral Laplace transform of $2\pi r\,\rho(|r|)$, 
\vspace{-.1mm}
and enters in the determination of the outside Yukawa potential 
\vspace{-.1mm}
induced by a new charge for a mediator of mass $m= k=$ $1/\lambda$. 
\,$\Phi(x)$ may be expressed as $\frac{3}{x^3}\,(x\,\cosh x - \sinh x) \times \bar\rho (x)/\rho_0$, where 
$\bar\rho(x)$ is an effective density decreasing (for $d\rho/dr <0)$  from the average $\rho_0$ at small $x$, down to the density $\rho(R)$  near the surface. An inversion formula allows one to recover the density distribution $\rho(r)$ from an analytic continuation of $\Phi(x)$, as 
$\rho(r) =\rho_0\, (2R/3\pi r)\int_0^\infty \Phi(ix) \, \sin (x\frac{r}{R})\ x\,dx\,$.

\vspace{2mm}

$\Phi(x)$ for the Earth is essential to determine limits on a very weak new force of finite range, as tested by  {\it MICROSCOPE\,},
depending on the density distribution within the Earth. Quite remarkably, much simplified density profiles,
 such as $\rho =  \rho_0\, 2R/3r$ or $\rho = $ $\rho_0\,(\frac54-\frac{r}{R}+ \frac{R}{3r})$,
 \vspace{-.4mm}
provide analytic expressions of $\Phi(x)$ and $\bar\rho(x)$ giving almost the same values as in a 5-shell model. 
$\Phi(x) \!=\! (\sinh\frac{x}{2} / \frac{x}{2})^2$
\vspace{-.4mm}
is valid to within $\simeq 1\,\%$ up to $x=4$.
$\Phi(x)= [\,7\,x^2 \cosh x-24 \,\cosh x+ 9\,x\sinh x -4\,x^2+24\,] / (4x^4)$
  is valid to within 1\,\%  for $x<\,64\,$ i.e.~$\lambda > $ 100 km or $m< 2\times 10^{-12}$ eV$/c^2$.
\,For $m=10^{-12}$\,eV/$c^2$ the coupling \hbox{limits} 
\vspace{.1mm}
are increased by $\simeq 34$ as compared to a massless mediator, to $|g_{B\!-\!L}|\! < 3.6 \times 10^{-24}$ and $|g_B| \!< 2.6 \times 10^{-23}$ for a spin-1 mediator, with slightly different limits in the spin-0 case.

\end{abstract}

\maketitle

\vspace{1mm}

\section{General overview}

\vspace{1mm}

The understanding of the Yukawa potential induced by an extended body, for an interaction of finite range, is essential to discuss its effects, independently of the particular nature of this interaction.

\vspace{2mm}

More specifically, new interactions may exist beyond strong, electromagnetic, weak and gravitational ones, and constitute elements of a new ``dark sector".  This is the case if the $SU(3)\times SU(2)\times U(1)$ symmetry group of the standard model is extended to an extra $U(1)$, often referred to as a ``dark $U(1)$".  Its spin-1 gauge boson $U$ may be massless or very light, and appears as a generalized dark photon mediating an extremely weak long-range interaction.
Its couplings to standard model particles generally involve a combination of the electric charge  with baryonic and leptonic numbers, possibly through their $B\!-\!L$ combination, in a grand-unified theory \cite{U,U2}. 

\vspace{2mm}

Such an interaction, which may also be induced by a spin-0 mediator \cite{dp,kw,dd}, could lead to apparent violations of the equivalence principle expressing the identity between inertial and gravitational masses. This principle, well tested by the E\"otv\"os experiment \cite{eot}, is at the basis of general relativity. Its expected limitations, especially when trying to include gravitational interactions in the quantum framework, and the search for new feeble interactions, have motivated further tests to a very high level of precision \cite{adel,ew}. \linebreak A very feeble interaction mediated by an extremely light $U$ boson may also be responsible for a spontaneous breaking of supersymmetry at a very large scale, associated with a huge vacuum energy density that may be responsible for the very rapid expansion of an inflationary Universe \cite{inf}.

\vspace{2mm}

Most notably, the {\it MICROSCOPE\,} experiment  \cite{microold,micronew} has monitored  the difference in accelerations between two Titanium and Platinum test masses freely-falling around the Earth in a drag-free satellite, orbiting at an average altitude $z\simeq 710$ km. It leads to an E\"otv\"os parameter $\,\delta_{\hbox{\scriptsize\,Ti-Pt}}=(-1.5\,\pm\, 2.3_{\text{\,stat}} \,\pm \,1.5_{\text{\,syst}})\,\times 10^{-15}$, i.e. at the 95\,\% confidence level

\vspace{-3mm}

\be
\left\{\ \ba{cccc}
\delta \! &< \,4.5\times 10^{-15}\ &\hbox{for}\ \delta>0\ ,
\vspace{1mm}\\
|\delta| \!&<\, 6.5\times 10^{-15}\ & \hbox{for}\ \delta <0\ .
\ea
\right.
\ee
The coupling limits are stronger for a coupling to $B\!-\!L$ or $L$ than for a coupling to $B$, and slightly different (by a factor $\sqrt{\,6.5/4.5}\,\simeq 1.2$) depending on the mediator spin \cite{fayetmicro,yuk}. The precise limits also involve the density profile of the Earth, and thus its hyperbolic form factor, obtained as the Laplace transform of its density distribution, related by analytic continuation to its ordinary form factor \cite{yuk}.

\subsection{General features for Yukawa potentials and form factors}

 If the new interaction considered is long-ranged i.e. for a range $\lambda\!=\! 1/m \gg R$ ($R$ being the radius of the sphere and $m=k$ the mediator mass, with $\hbar = c=\mu_0= \epsilon_0=1$), the induced potential at a distance $r>R\,$ from the center is very close to a Coulomb potential $Q/{4\pi r}$, $Q$ being the total new charge of the sphere.
In general the outside potential,  which would be $Q\,e^{-kr}/{4\pi r}$  for a pointlike sphere,
 gets increased by a factor $\Phi(x)$ where $x=kR=R/\lambda$, as the potential at a distance $r$ from the center may be 
 mostly generated, especially for shorter $\lambda$'s,  by parts of the sphere which are closer than its center, down to a minimum distance 
 $r-R$ (equal to the satellite altitude, $z\simeq 710$ km, in the case of {\it MICROSCOPE\,}).

\vspace{2mm}

The outside potential can then be expressed as a Yukawa potential
$ {\cal V}(r)=Q\ \frac{e^{-kr}}{4\pi r}\ \Phi(x)\,$,
where $\,\Phi(x\!=\!kR)\geq 1$ is the hyperbolic form factor of the density distribution $\rho(r)$. 
In the case of the Earth this one is taken to be the same as for the mass distribution.
This form factor  may be first defined as the bilateral Laplace transform 
of the density distribution $\rho(\vec r)$ in three dimensions, 
\vspace{-.9mm}
according to
$g(\vec k) = \langle \,e^{\vec k.\vec r}\,\rangle\!=\! \int \rho(\vec r)\ e^{\vec k.\vec r}\, d^3\vec r$, 
\,with $g(0) =1$ for  $\rho(\vec r)$ normalized to unity.  
\vspace{-.9mm}
If  the origin is a center of symmetry it may be rewritten as 
$ g(\vec k) = \langle \, \cosh\, {\vec k.\vec r}\ \rangle $.
\vspace{-.6mm}
For a spherically symmetric distribution
$g(\vec k)$ depends only on  $k = |\vec k|$ \,and  $\Phi(x\!=\!kR)$ may be identified with it, leading to  \cite{yuk}
\vspace{-4mm}

\be
\label{Phig}
\Phi(x) \,=\,g(k) =  \langle \, \cosh\, {\vec k.\vec r}\ \rangle = \langle \ \frac{\sinh kr}{kr}\ \rangle\ .
\ee
This hyperbolic form factor is related, by a duality ${\cal D}$ or by analytic continuation, 
\vspace{-.5mm}
to the ordinary form factor $f(k)=  \langle \, e^{i \vec k.\vec r}\, \rangle$, expressed as $\langle \,\frac{\sin kr}{kr}\,\rangle$, through the change $k\to \pm\, ik$, or $x^2 \to -\,x^2$, according to
\vspace{-2mm}

 \be
\label{d00}
\Phi(x)=  g(k) = \langle\  \frac{\sinh kr}{kr}\ \rangle \ \ \ \hbox{\large $ \stackrel{\cal D}{\longleftrightarrow}$}\ \ \
\tilde \Phi(x) =\Phi(ix)=  f(k) =\langle\  \frac{\sin kr}{kr}\ \rangle \ .
\ee
For an homogeneous sphere this reads
$\phi(x) = 3\, (x\cosh x-\sinh x)/x^3 \ \ \hbox{$ \stackrel{\cal D}{\longleftrightarrow}$}\ \
\tilde \phi(x) = 3\, (\sin x- x\,\cos x)/x^3\, .
$
The identity between the expressions (\ref{Phig}) of the form factors in $3d$ and using a radial coordinate may also be seen though their power series expansions in terms of $k$, or $x=kR$, using 
$\langle \, \cos^{2n}\theta \, \rangle = 1/(2n+1)$, so that
 \be
g(\vec k) = \langle \, \cosh\, {\vec k.\vec r}\ \rangle \ =\ \sum_0^\infty\  \frac{k^{2n}\,\langle \ (r\cos \theta)^{2n}\ \rangle}{(2n)\,!} \ =\ \sum_0^\infty \ \frac{k^{2n}\,\langle \,r^{2n}\,\rangle}{(2n+1)\,!}\,=\,\langle \ \frac{\sinh kr}{kr}\ \rangle = \,g(k)= \Phi(x) \  ,
 \ee
 and similarly for $f(k)=  \langle \, \cos\, {\vec k.\vec r}\ \rangle $, inserting the extra factor $(-1)^n$ in the power series expansion.

  \subsection{\boldmath An effective density $\bar\rho(x)$}

For each range $\lambda$ corresponding to $x= kR = R/\lambda$ we can define an effective density
$\bar\rho(x)$  such that the sphere generates, for a mediator of mass $m=k$, the same Yukawa potential as an homogenous sphere with density $\bar\rho(x)$.
$\Phi(x)$ may then be expressed as
\be
\Phi(x) = \frac{3}{x^3}\  (x\, \cosh x-\sinh x)\ \frac{\bar\rho(x)}{\rho_0}\ ,
\ee

\noindent
where $\rho_0$ is the average density. This may be rewritten as
\be
\label{rhobar00}
\bar\rho(x) =\,\rho_0 \ \,\frac{\Phi(x)}{\phi(x)} \,=\, 
\frac{\hbox{\small $\displaystyle\int_0^R$}\ \rho(r)\ rdr \, \sinh kr}
{\hbox{\small $\displaystyle\int_0^R$}\ rdr \, \sinh kr}\ \,.
\ee
We shall show in Sec.\,\ref{sec:dec} that this effective density $\bar\rho(x)$,  characteristic of the distribution of charges or masses within the sphere considered, decreases (for $d\rho(r)/dr <0$) from the average density $\rho_0$ at $x=0$ down to the density $\rho(R)$ near the surface of the sphere. For the Earth $\bar \rho(x)$ will be given in Table \ref{tab:comp} and represented in Fig.\,\ref{rhorho} (cf. Sec.\,\ref{sec:earth}, near the end of this paper).

\subsection{\boldmath From $2\pi r \rho(r)$ to $k g(k)$ through a bilateral Laplace transform, and back}

An interesting compact expression valid for both internal and external potentials, obtained in Sec. \ref{sec:yuk}, is 
\vspace{-6mm}

\be
{\cal V}(r) \,=\,\frac{Q}{2kr}\int_{-\infty }^\infty r'\rho(r')\ e^{-k|r-r'|}\, dr' \ ,
\ee
where the radial variable $r$ is allowed to have negative values, with ${\cal V}(r)$ and $\rho(r)$ extended as even functions of $r$.
We shall then see in  Sec.\,\ref{sec:inv} that $kg(k)$ appears as the bilateral Laplace transform of the odd function $2\pi r\,\rho(r)$, according to
\be
kg(k)= \int_{-\infty}^{\,\infty}\, 2\pi r\,\rho(r)\  e^{kr}\,dr\,=\,\int_0^\infty \rho(r)\ \frac{\sinh kr}{r}\ 4\pi r^2 dr \,=\, 
k \ \langle\,\frac{\sinh kr}{kr}\,\rangle\ .
\ee
An inversion formula allows one to recover the density $\rho(r)$ 
from the analytic continuation of the hyperbolic form factor $\Phi(x)= g(k)$, as
\vspace{-5mm}

\be
\label{inv}
r\rho(r)\,=\, \frac{1}{2\pi^2}\ \int_0^\infty kg(ik) \, \sin kr\,dk\,,
\ee
\vspace{-4mm}

\noindent
or equivalently in terms of $\Phi(ix)$,
\be
\label{inv0}
\rho(r)\,=\, \frac{1}{2\pi^2R^2 r}\, \int_0^\infty \Phi(ix) \, \sin\hbox{\large$\left(\right.$}x\frac{r}{R}\hbox{\large$\left.\right)$}\ x\,dx\,=\,\rho_0\ \frac{2R}{3\pi r}\,\int_0^\infty \Phi(ix) \, \sin\hbox{\large$\left(\right.$}x\frac{r}{R}\hbox{\large$\left.\right)$}\ x\,dx\ ,
\ee
with $\rho_0\,\frac{4\pi R^3}{3} = 1$.
This formula can be extended to apply, in the sense of distributions, to recover a pointlike distribution $\rho(\vec r )= \delta^3(\vec r)$ from its hyperbolic form factor $\Phi(x)=1$, with $\rho(r)$ for a pointlike distribution recovered as the non-conventional expression $\rho(r) = \delta(r)/(2\pi r^2)$, 
\vspace{.2mm}
correctly normalized to $\int _0^\infty \rho(r) \,4\pi r^2 dr = 2\,\int_0^\infty \delta (r) \,dr =1$.

\subsection{\boldmath A density distribution proportional to $1/r$}

We shall discuss in Sec.\,\ref{sec:hyp} expressions of $\Phi(x)$ and $\bar\rho(x)$ for various density distributions $\rho(r)$, including $\,\rho(r) = \rho_0 \ (p+3)\, r^p/(3\,R^p)$ proportional to $r^p$ for $r\leq R$, and even, most notably, $\rho(r)= \rho_0 \ 2R/3r$  proportional to $1/r$. This one, with a singularity at the origin partially regularized by the volume element $4\pi r^2 dr$, leads to an elegant expression of $\Phi(x)$,  developed in power series as follows:
\be
\label{phiinv}
\Phi(x) \,=\, \frac{2}{x^2}\,(\cosh x -1) = \left(\frac{\sinh {x}/{2}}{x/2}\right)^{\!2}=\,\sum_0^\infty\  \frac{2\,x^{2n}}{(2n+2)\,!}\, =  \ 1\, + \frac{x^2}{12}\,+ \frac{x^4}{360}\,+ \frac{x^6}{20\,160}\,+\frac{x^8}{1\,814\,400} \,+ \,...\ .
\ee
The $\,{x^2}/{12}\,$ term is associated with a moment of inertia $I \!= \frac23\,M\langle\, r^2\,\rangle = \frac13\,MR^2$, not  far from the \hbox{$.3308\,MR^2$ corresponding} to the Earth \cite{nasa}, and less than the $\frac25\,MR^2$ of an homogeneous sphere.
\,The effective density is
\vspace{-5mm}

\be
\bar\rho(x) =\, \rho_0\  \frac{2x\,(\cosh x-1)}{3\,(x\cosh x-\sinh x)}\ ,
\ee
decreasing down to $\rho(R) = \frac23\,\rho_0$ for large $x$ i.e. short ranges.
The ordinary form factor for this density distribution reads 
\be
\label{phiinv2}
\tilde \Phi(x)= \Phi(ix)  \,=\,\frac{2}{x^2}\,(1-\cos x) = \left(\frac{\sin {x}/{2}}{x/2}\right)^{\!2}=\,\sum_0^\infty\  \frac{2\,(-1)^n\,x^{2n}}{(2n+2)\,!}\, =  \, 1\, - \frac{x^2}{12}\,+ \frac{x^4}{360}\,- \frac{x^6}{20\,160}\,+ \,...\ .
\ee
It vanishes for $x = kR =2n\pi\,$ i.e. when the  radius $R$ is $n$ times the wavelength $2\pi/k$. This corresponds to an inside solution for the potential, obtained as
\be
{\cal V}_{\rm in} (r) =\, \rho_0\ \frac{2R}{3}\ \frac{\cos kr-1}{k^2 r}\ .
\ee
It vanishes at $r=0$ and, together with its derivative, at $r\!=\!R$ for $x= 2n\pi$, then corresponding  to a vanishing outside potential ${\cal V}_{\rm out} (r)=0$; i.e, when a time dependence is introduced, to a stationary wave with angular frequency $\omega = \pm\, k$, 
confined within the sphere.

\subsection{Application to the Earth}

We shall then consider the Earth in Sec.\,\ref{sec:earth}, with a moment of inertia $I\simeq .3308\,MR^2$. Its hyperbolic form factor $\Phi(x)$ has been evalued in \cite{yuk} in a simple 5-shell model inspired from \cite{prem}, distinguishing between inner and outer cores, inner and outer mantles, and crust, with successive radii taken as 1121, 3480, 5701, 6341 and 6371 km, respectively.
In fact the details of the internal structure of the core  are smoothed out for large $\lambda$'s and irrelevant for smaller ones. 
This will allow for the very simple expression $\,\Phi(x)\!=\! \frac{2}{x^2}\,(\cosh x -1)$, given in eq.\,(\ref{phiinv}) for the density profile $ \rho(r)= \rho_0 \ \frac{2R}{3r}$,  
\vspace{-.2mm}
with $I = \frac13\,MR^2$, 
to already provide, despite its singularity at the origin, a good approximation of the hyperbolic form factor, found to be valid to within  $\simeq 1\,\%$ up to $x =4$, and 5\,\% up to $x=10$, as compared with the 5-shell model. 

\vspace{2mm}

\begin{figure}
\caption{\ $\Phi(x)$ 
as a function of $x=R/\lambda$, for four density distributions: 
1) an homogeneous $\rho_0$, with $\phi(x) = $ $\frac{3}{x^3}\,(x\cosh x-\sinh x)$ (dotted green);
2) $\rho(r)=\rho_0 \,\frac{2R}{3r}$, with $\Phi(x) = \frac{2}{x^2}\,(\cosh x-1)$ (dotted red);
\vspace{-.5mm}
3) $\rho'(r)=$ $ \rho_0 \,(\frac54 - \frac{r}{R}+\frac{R}{3r})$,
with $\Phi'(x) =\frac{1}{4x^4}\,[\,7x^2 \cosh x-24 \cosh x+ 9x\sinh x -4x^2+24\,]$ (cf. Sec.\,\ref{sec:earth}, in blue);
\vspace{.1mm}
4) $\rho(r)$ in a 5-shell Earth model \hbox{(also in blue) \cite{yuk}.}
The last two curves are superposed, 
\vspace{.1mm}
coinciding to within $\simeq .5\,\%$ for  $x <50\,$.
$\,\Phi(x) = \frac{2}{x^2}\,(\cosh x-1)$, in dotted green,
slightly overestimates this result, by $\simeq 1\,\%$ for $x= 4$, and 4\,\% for $x = 8$.
\,$\phi(x)$
for an homogenous density, in dotted green, overestimates it 
by $\simeq 36\,\%$ for $x=8$.
\label{4phi}
\vspace{5mm}}
\includegraphics[width=8.5cm,height=6.2cm]{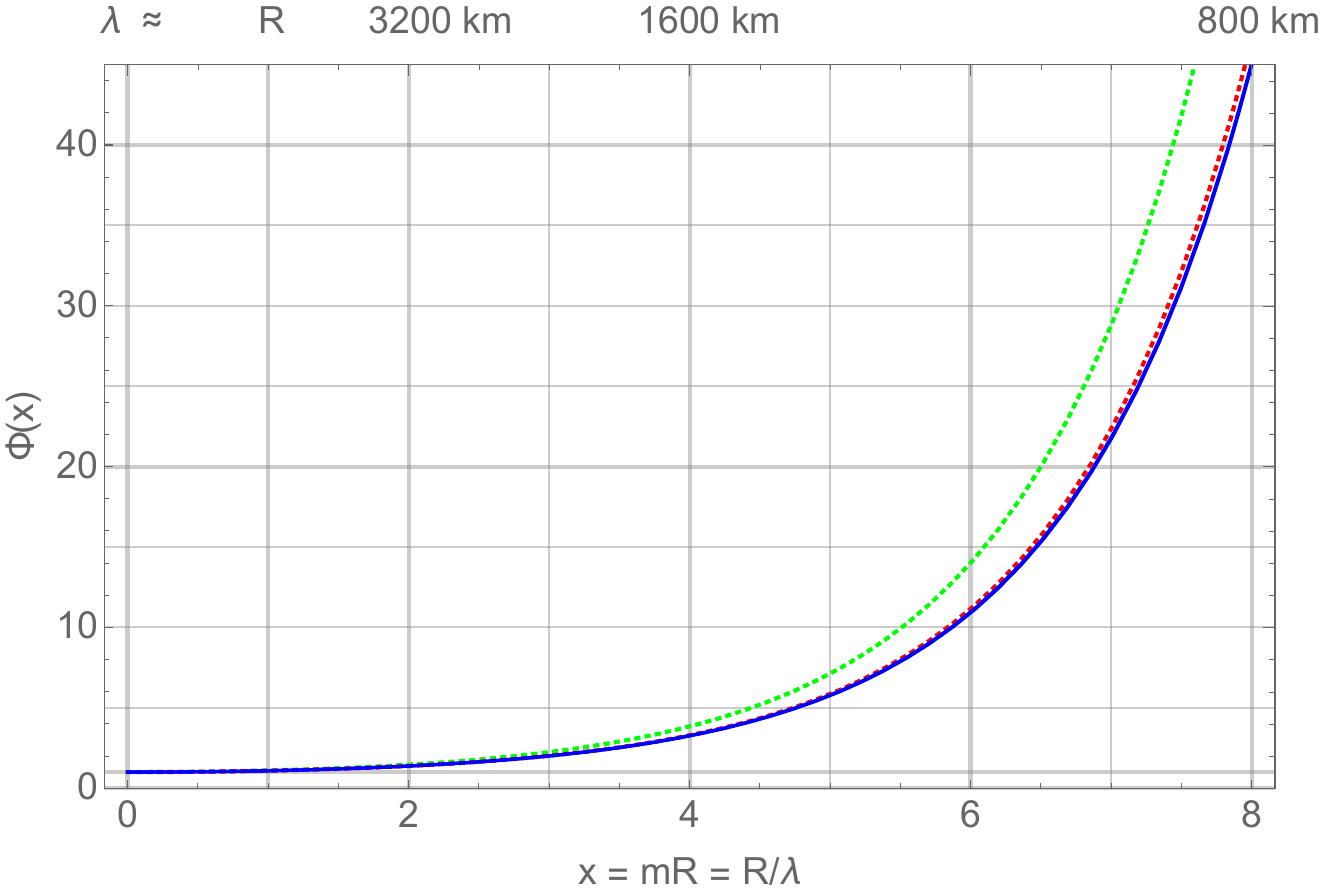}
\end{figure}

This is much better than the $\phi(x)= \frac{3}{x^3}\,  (x\, \cosh x-\sinh x)$ for an homogeneous sphere, 
\vspace{.2mm}
with $I=\frac25\,MR^2$. 
Comparisons are given in Fig.\,\ref{4phi} and Tables \ref{tab:comp},\,\ref{tab:r} in Secs.\,\ref{sec:hyp} and \ref{sec:earth}.\, 
\vspace{-.6mm}
The  power series expansion $\Phi(x) = 1 + \frac{x^2}{12}+ \,...$ in eq.\,(\ref{phiinv}) is thus also close to the $\Phi(x) = 1 + .0827 \,x^2 + \,...$ 
\vspace{.2mm}
obtained in a 5-shell model, recalled below in eq.\,(\ref{exp50}).

\vspace{2mm}
Even better, another simple expression of the density is obtained as the average between this density $\rho_0\,2R/3r$ and a density $\rho_l = \rho_0\,(\frac52-2\,\frac{r}{R})$, 
\vspace{-.4mm}
decreasing linearly from $\frac52\,\rho_0$ at the center down to $\frac12\,\rho_0$ near the surface, also with  $I=\frac13\,MR^2$. It  is thus equal to
\vspace{-3mm}

\be
\rho'(r)= \rho_0\ \left(\,\frac54-\frac{r}{R}+\frac{R}{3r}\,\right)\ ,
\ee
decreasing to $\frac{7}{12}\, \rho_0$ near the surface of the sphere, 
and again providing a moment of inertia $I=\frac13\,MR^2$.
Quite interestingly this density distribution provides almost exactly the same $\Phi(x)$ as in a 5-shell model of the Earth, to within $.7\,\%$ up to $x=64$, i.e. $\lambda$ down to 100  km (cf. Tables \ref{tab:comp}, \ref{tab:r} and Figs.\,\ref{4phi},\,\ref{rhorho} in Secs.\,\ref{sec:hyp} and \ref{sec:earth}). This occurs even if this smooth expression of $\rho(r)$ does not take into account a significant core/mantle density discontinuity around $R_c\simeq 3480$ km. 
\vspace{2mm}

This leads us to retain the following analytic expressions for $\Phi(x)$ and $\bar\rho(x)$, 
\be
\label{Phin0}
\left\{\ 
\ba{ccc}
\Phi'(x)\!&=&\!\dis\frac{1}{4x^4}\ [\,7x^2 \cosh x-24 \cosh x+ 9x\sinh x -4x^2+24\,]\ ,
\vspace{3mm}\\
\bar\rho'(x)\!&=&\! \dis \rho_0\ \,\frac{7x^2 \cosh x-24 \cosh x+ 9x\sinh x -4x^2+24}{12\,x\,(x \cosh x - \sinh x)}\ ,
\ea\right.
\ee
with $\bar\rho'(x)$ decreasing down to $\frac{7}{12}\,\rho_0$ at large $x$.
The power series expansion 
\be
\label{expPhin0}
\ba{ccl}
\Phi'(x)\! &=& \dis \,\sum_0^\infty \ \frac{7n^2+29n+24}{(2n+4)!} \ x^{2n}
=  1 + \frac{1}{12}\ x^2  + \frac{11}{4032} \ x^4+ \frac{29}{604\,800}\ x^6 + \frac{1}{1\,900\,800}\ x^8\,+\,...
\vspace{4mm}\\
&\simeq& 1+.0833\,x^2 + 2.73 \times 10^{-3}\,x^4+ 4.79 \times 10^{-5}\,x^6+ 5.26 \times 10^{-7}\,x^8+ 3.95 \times 10^{-9}\ x^{10}+\, ...\,, \!\!
\ea\!\!\!\!\!\!
\ee
is also very close to the corresponding expression in the 5-shell model \cite{yuk},
\be
\label{exp50}
\Phi_{\hbox{\footnotesize 5s}}(x) \simeq \,1 +  .0827 \,x^2 + 2.71 \times 10^{-3}\,  x^4 + 4.78\times 10^{-5}\ x^6 +  5.26 \times 10^{-7}\ x^8 + 3.95 \times 10^{-9}\ x^{10} +\,...\, .
\ee
\vspace{-1mm}

The two expansions (\ref{expPhin0}) and (\ref{exp50}) of $\Phi(x)$, although very similar,  cannot be the same everywhere, or the corresponding densities $\rho(r)$ would also be the same owing to the inversion formula (\ref{inv0}). Small differences, as for the $x^2$ and $x^4$ terms but which do not otherwise play a significant role, are thus mandatory.
Formulas (\ref{Phin0},\,\ref{expPhin0}) for $\Phi(x)$ and $\bar\rho(x)$ can thus be used as analytic substitutes for those obtained in the 5-shell model. It also appears unlikely that further adjustments of the density profile $\rho(r)$ by taking into account more shells can lead to significant modifications of the hyperbolic form factor $\Phi(x)$, and to the resulting limits on the couplings of a new interaction.

\subsection{Outline of the paper}

\vspace{-1mm}

Section \ref{sec:hyp2} deals with hyperbolic form factors 
\vspace{-.2mm}
expressed as $\Phi(x\!=\!kR)= g(k) = \langle \,\cosh\, \vec k.\vec r\,\rangle = \langle \,\frac{\sinh kr}{kr}\,\rangle$, and their relations to ordinary form factors through a duality acting as $k\to\pm\, ik$ or $x^2\to - x^2$, 
\vspace{.1mm}
exchanging $\cosh x$ and ${\sinh x}/{x}$ with  $\cos x $ and ${\sin x}/{x}$. 
\vspace{-.6mm}
\,Section \ref{sec:yuk} discusses the inside and outside potentials for a sphere, the latter expressed as 
$Q\ \frac{e^{-kr}}{4\pi r}\times \Phi(x)$, also extending the radial variable $r$ to negative values.
Section \ref{sec:exp} discusses expansions of the form factors in terms of $x=kR$, with coefficients fixed by the moments $\langle \,r^{2n}\,\rangle$.
\,Section \ref{sec:dec} shows that the effective density $\bar\rho(x)= \rho_0\,\Phi(x)/\phi(x)$ is a decreasing function of $x$, for $d\rho(r)/dr <0$, down to the density $\rho(R)$ near surface.

\vspace{2mm}

Section \ref{sec:hyp} discusses form factors for various density distributions, including a density proportional to $1/r$, and more generally to $r^p$, for $r\!\leq \!R$.
\,Section \ref{sec:inv} shows that $kg(k)$ appears as the bilateral Laplace transform of $2\pi r\, \rho(|r|)$, and how an inversion formula allows one to recover $\rho(r)$ from 
$\Phi(ix)$.
\,Section \ref{sec:earth} deals with the Earth, 
showing how very simple density profiles can lead to surprisingly good analytic expressions of 
$\Phi(x)$ and $\bar\rho(x)$, such as $\Phi(x)= (\sinh \frac{x}{2}/\frac{x}{2})^2$, or the more elaborate expressions in eqs.\,(\ref{Phin0}).

\section{Hyperbolic form factors}
\label{sec:hyp2}

\subsection{Hyperbolic form factors in three dimensions}

The form factor for a density distribution $\rho(\vec r)$ is taken 
\vspace{-.8mm}
as its Fourier transform,
$\,f(\vec k) = \int \rho(\vec r)\ e^{i\vec k .\vec r}\ d^3\vec r$ $=\,\langle\,e^{i\,\vec k . \vec r}\, \rangle$\,,
with
$f(0)=  1$ for a distribution normalized to unity. We define in a similar way an hyperbolic form factor as the bilateral Laplace transform of $\rho(\vec r)$,
\be
\label{g}
g(\vec k) = \int\,\rho(\vec r)\ e^{\,\vec k .\vec r}\ d^3\vec r\,\,=\,\langle\,e^{\vec k . \vec r}\, \rangle\,,
\ee
with $g(0)= 1$, 
\vspace{-.7mm}
provided it is well defined as for $\rho(\vec r)$ with compact support, or integrable and decreasing sufficiently rapidly at infinity. 
\vspace{-.4mm}
$g(\vec k)$ also appears as given by an analytic extension of $f(\vec k)$, through
$g(\vec k)=f(-i\,\vec k)$.
If we translate $\rho(\vec r)$ according to $\rho(\vec r)\to \rho'(\vec r) = \rho(\vec r -\vec a$),
\vspace{-.5mm}
$g(\vec k)$ is rescaled according to
$ g(\vec k)\to g'(\vec k)= e^{\,- \vec k.\vec a}\, g(\vec k)$.
These form factors are related by a  duality transformation $\cal D$ squaring to the identity, such that
\vspace{-4mm}

\be
\label{d0}
g(\vec k) = \int\rho(\vec r)\ e^{\,\vec k .\vec r}\ d^3\vec r\ \ \ \ \hbox{\large $\stackrel{\cal D}{\longleftrightarrow}$}\ \ \ 
f(\vec k) = \int \rho(\vec r)\ e^{i\vec k .\vec r}\ d^3\vec r= \,g(i\vec k)\ ,
\ee
exchanging the bilateral Laplace and Fourier transforms of  $\rho(\vec r)$. 
If the origin is a center of symmetry eq.\,(\ref{g}) 
\vspace*{-2mm}
may be rewritten as
\be
\label{d0bis}
g(\vec k) \,=\,\langle\ \cosh\,{\vec k . \vec r}\ \rangle\,.
\ee
$\rho(r)$ may then be recovered 
\vspace{-.8mm}
from the inverse Fourier transform of $f(\vec k)$, as
$\rho(\vec r\,)=\frac{1}{2\pi}\int\! f(\vec k)\,  e^{-i\vec k .\vec r}\, d^3\vec k\,$.
\vspace{-.2mm}
The inverse Laplace transform of $g(\vec k)$ may then be expressed as
\vspace{-3mm}

\be
\label{invl}
\rho(\vec r\,) = \frac{1}{2\pi} \int \,g(i\vec k)\,  e^{-i\vec k .\vec r}\ d^3\vec k\ .
\ee

\vspace{1mm}

For a parallelepiped of center $O$ and sides $2a,\,2b,\,2c$, we have
\be
\left\{\ \ba{ccccl}
f(\vec k)&=&\frac{1}{8\,abc}\  \hbox{$\int_{-a}^a\int_{-b}^b\int_{-c}^c$}\ \dis e^{\,i \vec k.\vec r}\ d^3\vec r\, &=&\dis \frac{\sin k_x a}{k_x a}\ \frac{\sin k_y b}{k_y b}\ \frac{\sin k_z c}{k_z c}\ ,
\vspace{2mm}\\
g(\vec k)&=&\frac{1}{8\,abc}\,\hbox{$\int_{-a}^a\int_{-b}^b\int_{-c}^c$}\ \dis e^{\,\vec k.\vec r}\ d^3\vec r\, &=&\dis \frac{\sinh k_x a}{k_x a}\ \frac{\sinh k_y b}{k_y b}\ \frac{\sinh k_z c}{k_z c} \,=\, f(-i\vec k)\ ,
\ea\right.
\ee
\vspace{1mm}

\noindent
and for a gaussian ellipsoïd with 
$\,\rho(r) =$ \hbox{\small $ \dis \frac{1}{abc\,(2\pi)^{3/2}}$}\,
$e^{-\,\hbox{$(\frac{x^2}{2a^2}+\frac{y^2}{2b^2}+\frac{z^2}{2c^2})$}}$,
\be
\label{ge2}
g(\vec k)\,=\,e^{\,\hbox{$(\frac{k_x^2a^2}{2}+\frac{k_y^2b^2}{2}+\frac{k_z^2c^2}{2})$}}
\ \ \ \ \hbox{\large $\stackrel{\cal D}{\longleftrightarrow}$}\ \ \ \
f(\vec k)\,=\,e^{-\,\hbox{$(\frac{k_x^2a^2}{2}+\frac{k_y^2b^2}{2}+\frac{k_z^2c^2}{2})$}}\ .
\ee

\subsection{The hyperbolic form factor of a sphere}

For a spherically symmetric density distribution normalized to unity, one has
\be
\label{gk}
\left\{\
\ba{ccc}
f(\vec k)= f(k) =\,\int_0^\infty\!\rho(r)\int_0^\pi e^{ikr\cos\theta}\ 2\pi\sin\theta\,d\theta\  r^2 dr\ ,
\vspace{3mm}\\
g(\vec k) =g(k) = \,\int_0^\infty\!\rho(r)\int_0^\pi \,e^{\,kr\cos\theta}\ 2\pi\sin\theta\,d\theta\  r^2 dr\ ,
\ea
\right.
\ee
so that
\be
\label{fg}
\left\{\
\ba{ccccccccccc}
f(k)\ \ \ \ \ &=& \dis \int_0^\infty \rho(r)\ \frac{\sin kr}{kr}\ 4\pi r^2dr  &=&\dis \langle\  \frac{\sin kr}{kr}\ \rangle\,,
\vspace{2mm}\\
g(k)\, =\,f(\mp\,i k) \!&=&\dis \int_0^\infty \rho(r)\ \frac{\sinh kr}{kr}\ 4\pi r^2dr &=&\dis\langle\  \frac{\sinh kr}{kr}\ \rangle \,.
\ea
\right.
\ee
They verify the duality relations (\ref{d0}), now associated with the change $\,k\, \leftrightarrow \pm\, ik\,$, 
\be
\label{d}
\framebox [8cm]{\rule[-.3cm]{0cm}{.9cm} $ \dis
g(k) = \langle\  \frac{\sinh kr}{kr}\ \rangle \ \ \ \hbox{\large $ \stackrel{\cal D}{\longleftrightarrow}$}\ \ \ f(k) =\langle\  \frac{\sin kr}{kr}\ \rangle \ .
$}
\ee

\vspace{1mm}

For an homogeneous sphere of radius $R$, one has, with $x = kR$, 
\be
\label{hom}
g(k) = \phi(x)  = \frac{3}{x^3}\ (\cosh x - x\, \sinh x) 
\ \ \ \ \hbox{\large $\stackrel{\cal D}{\longleftrightarrow}$}\ \ \ \
f(k)=\tilde \phi(x) = \dis \frac{3}{x^3}\ (x\,\sin x - \cos x)\ .
\ee
For a spherically symmetric gaussian distribution, one has from eq.\,(\ref{ge2}), with $a\!=\!b\!=\!c\!=\!\sigma$,
\be
g(k) = \,e^{\hbox{\,$\frac{k^2\sigma^2}{2}$}}\ \ \ \ \hbox{\large $\stackrel{\cal D}{\longleftrightarrow}$}\ \ \  f(k) = \,e^{-\,{\hbox{$\frac{k^2\sigma^2}{2}$}}}\,,
\ee
as also obtained in spherical coordinates from  eq.\,(\ref{fg}), with
\be
\ba{ccccl}
g(k)  &=&  \hbox{\normalsize $\dis \frac{1}{(\sigma\, \sqrt{2\pi})^3}\, \int_0^\infty \!  e^{\hbox{\footnotesize$\dis -\frac{r^2}{2\sigma^2}$}}\ \frac{\sinh kr}{k}\ 4\pi rdr$}
 &=&  \hbox{\normalsize $\dis \sqrt \frac{2}{\pi}\, \int_0^\infty \! e^{\hbox{\,--\,$ \frac{u^2}{2}$}}\,\ \frac{\sinh\! \,k\sigma u}{k\sigma}\ \,u du $}
\vspace{2.8mm}\\ &=& 
 \hbox{\normalsize $\dis  \sqrt \frac{2}{\pi}\ \int_0^\infty \! e^{\hbox{\,--\,$ \frac{u^2}{2}$}} \,\cosh\! {\,k\sigma u}\ \,du $}
 &=& \hbox{\normalsize $\dis  \frac{1}{\sqrt{2\pi}}\ \dis \int_{-\infty}^\infty  e^{\hbox{\,--\,$ \frac{u^2}{2}$}}\,e^{\,k\sigma u}\ du $}
\dis \ =\ e^{\hbox{\,$\frac{k^2\sigma^2}{2}$}}  \,.
\ea
\ee

\vspace{0mm}

\section{The inside and outside potentials of a sphere}
\label{sec:yuk}

\subsection{Expression of the potential}

We consider in three space dimensions an interaction of range $\lambda =1/k$, that may be given by $1/m$ where $m=k$ is the mediator mass (with $\hbar \!= \! c\!= \! \mu_0\!= \! \epsilon_0 \!= \! 1$).
The potential ${\cal V}(\vec r\,)$ induced by the new charge distribution $Q\,\rho(\vec r)$ obeys the Poisson-like equation  $(-\triangle + k^2)\, {\cal V}(\vec r\,) = Q\,\rho(\vec r)$ and is given by
\be
{\cal V}(\vec r\,)\,=\,Q\, \int \rho(\vec r\,') \ \,\frac{e^{-k |\vec r-\vec r\,'| } }{4\pi\,|\vec r-\vec r\,' |}\ d^3\vec r\,'\,.
\ee
For a spherically symmetric distribution of center $O$, it can be evaluated with $\,l^2 \!= \!|\vec r-\vec r\,'|^2=r^2\!+\!r'^2 -2 r r'\cos \theta$,
and $ {\sin\theta\, d\theta}/{l}={dl}/{rr'}$,
so that 
\be
\label{V0}
\ba{ccl}
{\cal V}(r)\!&=&  Q \,\hbox{\small$\dis \int_0^\infty \!\!\rho(r\,') \int_0^\pi \,\frac{e^{-k l } }{4\pi\,l}\ \,2\pi \sin\theta\,d\theta\ r'^2\,dr'$}
=\,Q \, \hbox{\small$\dis \int_0^\infty \!\!\rho(r\,')\  \frac{r'dr'}{2r}\,  \int_{|r-r']}^{r+r'} \,e^{-k l }\ dl $}\,,
\vspace{3mm}\\
\!&=&\dis\ \frac{Q}{4\pi r}\, \int_0^\infty\! \rho(r')\ \,\frac{e^{-k|r-r'|}- e^{-k(r+r')}}{2kr'}\ \,4\pi r'^2dr\,'\ .
\ea
\ee
\vspace{1mm}

One can identify the contributions from the internal shells (with $r'<r$) and the external ones (with $r'>r$).
\,The elementary potential induced by a thin shell of radius $r'$ and elementary charge $Q\,\rho(r')\,4\pi r'^2 dr'$
reads
\be
\label{thin}
\ba{ccl}
d\,{\cal V}\,(r)&=&\dis \frac{Q\,\rho(r')\,4\pi r'^2 dr'}{4\pi r}\, \ \frac{e^{-k |r-r'|}- e^{-k(r+r')}}{2kr'}
\vspace{3mm}\\
&=& \dis\ Q\ [\rho(r')\,4\pi r'^2 dr']\, \times 
\left\{\ \ba{ccccc} \dis \frac{e^{-kr}}{4\pi r}\ \frac{\sinh kr'}{kr'}\ \ & \hbox{outside the shell}\ (\,r>r'\,)\ ,\vspace{2mm}\\
\dis \frac{e^{-kr'}}{4\pi r'}\ \frac{\sinh kr}{kr}\ \ & \hbox{inside the shell}\ (\,r<r'\,)\ .
\ea \right.
\ea
\ee
For a finite radial extension $R\,$ the outside potential is obtained from eq.\,(\ref{V0}) as
\be
\label{Vout}
{\cal V}_{\rm out }(r)\,=\,Q\ \frac{e^{-kr}}{4\pi r}\, \int_0^R\! \rho(r')\ \frac{\sinh kr'}{kr'}\ 4\pi r'^2dr\,'\,=\ Q \ \frac{e^{-kr}}{4\pi r}\ \langle\ \frac{\sinh kr'}{kr'}\ \rangle \ .
\ee
It is the Yukawa potential of a pointlike charge, $Q \, {e^{-kr}}/{4\pi r}$, multiplied by the hyper\-bolic form factor
\be
\label{gk2}
\framebox [5.9cm]{\rule[-.3cm]{0cm}{.85cm} $ \dis
\Phi(x\!=\!kR) = g(k) = \,\langle\ \frac{\sinh kr}{kr}\ \rangle \ ,
$}
\ee
given in eqs.\,(\ref{fg},\,\ref{d}). For $k\!=\!0$ corresponding to a massless mediator $\Phi=1$ and we recover a Coulomb potential $Q/(4\pi r)$, outside the sphere of radius $R$.

\vspace{2mm}

A spherically symmetric distribution $Q\,\rho(r)$ generates the same outside Yukawa potential as a pointlike charge $Q$ at its center, multiplied by the hyperbolic form factor $\Phi(x)$. \,The interaction potential, and force, between two separated spheres with centers at a distance $r\!> R_1+R_2\, $  are also the same as for two pointlike charges at their centers, multiplied by  $\Phi_1(kR_1)\,\Phi_2(kR_2)$.

\subsection{\boldmath A compact expression for the potential, with $r$ extended to negative values}
\label{expv}

The potential ${\cal V}(r)$ in eq.\,(\ref{V0}) may be recovered from eq.\,(\ref{thin}) as a sum of the potentials induced by inner and outer spherical shells of radii $r'$, as 
\be
\!{\cal V}(r) \,=\,\dis Q\,\left[\,\frac{e^{-kr}}{r}\int_0^r \rho(r')\ \frac{\sinh kr'}{kr'}\ r'^2\,dr'\,+\, \frac{\sinh kr}{r} \int_r^\infty \rho(r')\ \frac{e^{-kr'}}{kr'}\ r'^2\,dr'\,\right]\ .
\ee
This may be elegantly reexpressed 
\vspace{-.1mm}
by extending $\rho(r)$ to an even function of $r$  through $\rho(-r)=\rho(r)= \rho(|r|)$, \,still normalized to unity through 
$\int_{-\infty}^{\, \infty}\,  \rho(r)\,2\pi r^2dr=\int_0^\infty \rho(r)\,4\pi r^2dr=1\,$,
so that
\be
{\cal V}(r) = \frac{Q}{2kr}\left[\,\int_{-r}^r\! r'\rho(r')\ e^{-k(r-r')}\, dr' + \! \int_r^\infty \!r'\rho(r')\ e^{-k(r'-r)}\, dr' +\! \int_{-\infty}^{-r}\! r'\rho(r')\ e^{-k(r-r')}\,dr'\,\right].
\ee
This provides a general formula for both internal and external potentials, and even if $\rho(r)$ extends up to infinity:
\vspace{-4mm}

\be
\label{vmr0}
\framebox [6.8cm]{\rule[-.35cm]{0cm}{.95cm} $ \dis
{\cal V}(r) \,=\,\frac{Q}{2kr}\int_{-\infty }^\infty r'\rho(r')\ e^{-k|r-r'|}\, dr' \ .
$}
\ee
It is equivalent to (\ref{V0}) but with an integration on $r$ from $-\,\infty$ to $\infty$, or $-R$ to $R$ for a finite radial exten\-sion $R$.
Formula (\ref{V0}) expresses how the effect of the density  $\rho(r')$ at radius $r'>0$ propagates into the  value of the potential ${\cal V}(r)$ at radius $r>0$, with a propagation in both directions involving a reflection at the origin.
This is automatically taken into account through the consideration of negative values of $r'$ in eq.\,(\ref{vmr0}) as corresponding to the virtual images of the ``physical'' points at $r'>0$, images which also act as a source of the potential ${\cal V}(r)$ at a positive radius $r$.

\vspace{2mm}

For a sphere of radius $R$ the outside potential  (\ref{vmr0}) reduces, with $e^{-k|r-r'|}= e^{-kr}\,e^{kr'}$, to the Yukawa potential in eq.\,(\ref{Vout}), recovered as
\be
\ba{c}
{\cal V}_{\rm out}(r)\,=\,Q\ 
\dis \frac{e^{-kr}}{4\pi r}\int_{-R}^R 4\pi \rho(r')\ \frac{e^{kr'}}{2kr'}\, r'^2 \,dr' \,=\,
 Q\  \frac{e^{-kr}}{4\pi r} \ \langle\, \frac{\sinh kr'}{kr'}\, \rangle
\, .
\ea
\ee
Expression (\ref{vmr0}) of ${\cal V}(r)$ allows for an immediate verification of the Poisson-like equation for ${\cal V}(r)$, with a smooth treatment of the otherwise singular point at $r'=0$ thanks to the consideration of negative values of $r'$, 
\be
\ba{ccl}
(-\triangle+k^2)\, {\cal V}(r) = \hbox{\small$\dis \frac{1}{r}$} \left(\hbox{\small$\dis -\frac{\partial^2\ }{\partial r^2}$}\!+k^2\right) r{\cal V}(r) \!&=&\!\dis \frac{Q}{2kr}
\int_{-\infty }^\infty \!r'\rho(r')\, \underbrace{\left[ \left(\hbox{\normalsize $\dis -\frac{\partial^2\ }{\partial r^2}$}+k^2\right)  e^{-k|r-r'|}\right]}_{\hbox{\normalsize$\dis 2k\, \delta(r-r')$}}\, dr' 
\vspace{-1mm}\\
 &=&Q\ \rho(r)\ .
\ea
\ee

\subsection{More on the duality between hyperbolic and ordinary  form factors}
\vspace{-1mm}

The duality relations (\ref{d0},\,\ref{d}) between hyperbolic and ordinary form factors may be further understood from the
change, in the Yukawa potential ${\cal V}_{\rm out}(r)$, of  ${e^{-kr}}/{r} $ into $ {e^{ikr}}/{r}$ through $\,k \rightarrow - \,ik$\,, 
so that
\be
\frac{e^{-kr}}{r}\ \ \ \stackrel{\cal D}{\longleftrightarrow}\ \ \ \frac{e^{ikr}}{r}\ .
\ee
The Poisson-like equation for the potential ${\cal V}(r)$ induced by a mediator of mass $m=k$ is changed into a Poisson equation for a time-dependent massless field,
\be
\label{poiss2}
(\,-\,\triangle + k^2\,)\, {\cal V}(r)=Q\,\rho(r)\ \ \ \stackrel{\cal D}{\longleftrightarrow}\ \ \ (\,-\,\triangle -\,k^2\,)\, {\cal V}(r)=Q\,\rho(r)\ .
\ee
The latter equation, written as $\,\Box  \,{\cal V}(r)\,e^{-\,i\omega t}\!=Q\,\rho(r)\,e^{-\, i\omega t}$, is the same as for time-dependent potential and charge distributions, taken as the real parts of fields oscillating with time at angular frequencies $\omega = \pm \,k$ (i.e. proportional to $\cos\, (kt +\varphi))$.
The outside Yukawa potential for a  static field of mass $m=k$, expressed as 
\vspace{-6mm}

\be
\label{calv1ter}
{\cal V}_{\rm out}(r)\,=\,Q\,
 \int \rho(r')\ d^3\vec r\,' \ \,\frac{e^{-k |\,\vec r-\vec r\,'\,|}}{4\pi\,|\vec r-\vec r\,'|}\,=
 \ Q\  \langle \,\frac{e^{-k |\,\vec r-\vec r\,'\,|}}{4\pi\,|\,\vec r-\vec r\,'\,|}\,\rangle\,=\,
 Q\ \frac{e^{-kr}}{4\pi r}\,\ \langle \ \frac{\sinh kr}{kr}\ \rangle
\, ,
\ee
becomes
\vspace{-5mm}

\be
\label{calv1quat}
{\cal V}_{\rm out}(r)\,=\,Q\,
 \int \rho(r)\ d^3\vec r\,' \,\ \frac{e^{\, ik |\,\vec r-\vec r\,'\,|}}{4\pi\,|\,\vec r-\vec r\,'\,|}
  \,=\ Q\  \langle \,\frac{e^{\, ik |\,\vec r-\vec r\,'\,|}}{4\pi\,|\,\vec r-\vec r\,'\,|}\,\rangle
  \,=\,Q\ \frac{e^{\, ikr}}{4\pi r} \ \,\langle \, \frac{\sin kr}{kr}\,\rangle   
  \ .
\ee
displaying the correspondence  (\ref{d}) between hyperbolic and ordinary form factors.

\vspace{1mm}

\section{\!\!\boldmath Expanding form factors in terms of the even moments of $\rho(r)$}
\label{sec:exp}

The form factors  in eqs.\,(\ref{fg}) may be expanded in terms of the even moments of the density distribution, as
\vspace{-6mm}

\be
\label{ex1}
\left\{ \ba{ccccc}
g(k)\! &=&\dis \sum_0^\infty\ \frac{1}{(2n+1)!}\,\int_0^\infty \rho(r)\,(kr)^{2n}\,4\pi r^2dr
&=&\dis \sum_0^\infty\ \frac{1}{(2n+1)!}\ \frac{\langle\ r^{2n}\rangle}{\lambda^{2n}}\, ,
\vspace{2mm}\\
f(k)= g(ik)\! &=&\dis \sum\ \frac{(-1)^n}{(2n+1)!} \,\int_0^\infty\rho(r)\, (kr)^{2n}\,4\pi r^2dr
&=&\dis \sum_0^\infty\ (-1)^{n}\, \frac{1}{(2n+1)!}\, \frac{\langle\ r^{2n}\rangle}{\lambda^{2n}}\,,
\ea \right. \!\!
\ee
with $\lambda = 1/k$. This $\lambda$ is, for $g(k)$, the range of the Yukawa interaction considered, 
and for $f(k)$, $\lambda =1/k= \lambda' / 2\pi$ where $\lambda'= 2\pi\lambda $ is a wavelength\,\,\footnote{It could have been more judicious to define instead 
the range of a Yukawa interaction as $\lambda \!\!\! / = \lambda/2\pi = 1 /k$, 
corresponding for $f(k)$ to spherical waves $\propto e^{i kr}/4\pi r $ with wavelengths $\lambda = 2\pi/k = 2\pi \lambda\!\!\! / \, $.
}.
\,This also reads
\be
\label{ex2}
\left\{\ \ba{ccccccc}
\Phi(x) = g(k) &=&\dis \langle\ \frac{\sinh kr}{kr}\ \rangle 
&=&\dis \sum_0^\infty\ \frac{x^{2n}}{(2n+1)!}\ \frac{\langle\ r^{2n}\rangle}{R^{2n}}&=&
\dis\sum_0^\infty \ \frac{A_{2n}}{(2n\!+\!1)!}\ \,x^{2n}\, ,
\vspace{2mm}\\
\tilde \Phi(x)= f(k)&=&\dis \langle\ \frac{\sin kr}{kr}\ \rangle 
\!&=&\!\dis \sum_0^\infty \,(-1)^{n}\,\frac{x^{2n}}{(2n+1)!}\ \frac{\langle\ r^{2n}\rangle}{R^{2n}}&=&
\dis\sum_0^\infty  \, (-1)^n\ \frac{A_{2n}}{(2n\!+\!1)!}\ \,x^{2n} ,
\ea \right.
\ee
$R\,$ being the radius of the sphere, or some typical radius associated with $\rho(r)$. 
\,For a distribution $\rho(r)$ normalized to unity, with $\rho_0\times 4\pi R^3/3= 1$, 
one has
\be
\langle\  r^{2n}\ \rangle\ = \int_0^\infty\, \frac{\rho(r)}{\rho_0}\ \frac{3\,r^{2n+2}\, dr}{R^{3}}\ = \ A_{2n}\ R^{2n}\ .
\ee

The same results are obtained by expanding the 3d expressions of $g(\vec k)$ and $f(\vec k)$, with 
$\vec k.\vec r= kr\,\cos\theta$ and integrating on $\theta$ using $\langle\ (\cos\theta)^{2n}\,\rangle = 1/(2n+1)$
 so that we recover the above expansions (\ref{ex2}), now obtained as
 \be
\label{ex20}
\left\{\ \ba{ccccccc}
g(k) \!&=&\!\dis \langle\ \cosh \, \vec k .\vec r\ \rangle 
\!&=&\dis \sum_0^\infty\ \frac{x^{2n}}{(2n+1)!}\ \frac{\langle\ r^{2n}\rangle}{R^{2n}}&=&
\dis\sum_0^\infty \ \frac{A_{2n}}{(2n\!+\!1)!}\ \,x^{2n}\, ,
\vspace{2mm}\\
f(k)\!&=&\!\dis \langle\ \cos\, \vec k .\vec r\ \rangle 
\!&=& \dis \sum_0^\infty \ \,(-1)^{n}\,\frac{x^{2n}}{(2n+1)!}\ \frac{\langle\ r^{2n}\rangle}{R^{2n}}&=&
\dis\sum_0^\infty  \, (-1)^n\ \frac{A_{2n}}{(2n\!+\!1)!}\ \,x^{2n} .
\ea \right.
\ee

\vspace{2mm}

All this allows us to view $\Phi(x)$ and $\tilde \Phi(x)$ as holomorphic functions of $x =kR$, on the appropriate domain for distributions $\rho(r)$ extending up to infinity. With $\langle\, r^{2n}\,\rangle \leq R^{2n}$ for a distribution with a finite radial extension $R$ these expansions are absolutely convergent for all $x$. They read
 \be
 \label{hol0}
 \framebox [15cm]{\rule[-1.05cm]{0cm}{2.3cm} $ \dis
 \left\{\ \ba{ccccccc}
 \Phi(x)&=& \langle\ \cosh \, \vec k .\vec r\ \rangle &=& \dis \langle \, \frac{\sinh kr}{kr}\, \rangle  &=&\dis  
 1\, +\, \frac{\langle\, r^2\,\rangle}{6 \,R^2}\ x^2\, +\,\frac{\langle\, r^4\,\rangle}{120\,R^4}\  x^4\,    +\,\frac{\langle \,r^6 \,\rangle}{5\,040\,R^6} \ x^6\,   +\,...\ ,
 \vspace{3mm}\\
 \tilde \Phi(x)&=& \langle\ \cos \, \vec k .\vec r\ \rangle &=&\dis  \langle \, \frac{\sin kr}{kr}\, \rangle  &=&\dis 
 1\, -\, \frac{\langle\, r^2\,\rangle}{6 \,R^2}\ x^2\, +\,\frac{\langle\, r^4\,\rangle}{120\,R^4}\  x^4\,    -\,\frac{\langle \,r^6 \,\rangle}{5\,040\,R^6} \ x^6\,   +\,...\ .
 \ea\right.
 $}
 \ee
 With
 \vspace{-7mm}
 
 \be
 \Phi(x) \ \ \ \hbox{\large $\stackrel{\cal D}{\longleftrightarrow}$}\ \ \ \tilde\Phi(x) = \Phi(ix)\ 
 \ee
the hyperbolic and ordinary form factors are related by the duality transformation $\cal D$ according to which $x^2 \leftrightarrow -\,x^2$, each one also provided by an analytic continuation of the other.

\vspace{2mm}

\begin{table}
\caption{\ Relations between the density $\rho(r)$ and its moments $\langle \,r^{2n}\,\rangle$, 
\vspace{-.5mm}
the hyperbolic form factor 
$g(k)= $ $\Phi(x\!=\!kR) = \langle\,\frac{\sinh kr}{kr}\,\rangle$
and its dual, $f(k)\!= \Phi(ix)$ $=\langle\,\frac{\sin kr}{kr}\,\rangle $, 
also appearing as the form factor
$\langle \,e^{i\vec k.\vec r}\,\rangle$\,.
 \,For a density normalized to unity, $kg(k) \!= \int_0^\infty 4\pi r\, \rho(r)\, \sinh kr\, dr$ and 
 $kf(k)\! = \int_0^\infty 4\pi r\, \rho(r)\, \sin kr\, dr$
 appear as 
the bilateral Laplace and Fourier transforms of $2\pi r\rho(|r|)$, respec\-tively.
 The density  may be recovered as 
 $\rho(r)  =$ $\rho_0\ 2R/(3\pi r)\, \int_0^\infty \Phi(ix)\,  \sin (x\frac{r}{R})\ x\,dx$, 
as seen later in Section \ref{sec:inv}.
 \vspace{.3mm}
\vspace{1mm} \\ }
\label{relations}
 \hbox{\small $\ba{c}
$\framebox [15.8cm]{\rule[-2cm]{0cm}{4.35cm} $ \dis \hspace{12mm}
\ \ \ \ba{ccc}
&&\hspace{-18mm}\dis \underline{\hbox{\it hyperbolic form factor}}:\  \ \hspace{4mm}
g(k) = \Phi(x) =  \, \langle \, \cosh \vec k.\vec r\, \rangle\,=\, \langle \, \frac{\sinh kr}{kr}\, \rangle\,=\, \dis \sum_0^\infty\ \frac{\langle \,r^{2n}\rangle}{(2n+1)! \ R^{2n}}\ \,x^{2n}\,
\vspace{2mm}
\\
&\hspace{-10mm}
\swarrow\!\!\!\!\!\!\nearrow\  \hbox{(Laplace)}
\vspace{2mm}\\
\hspace{-36mm}\rho(r)  =\rho_0\, \hbox{\normalsize $\frac{2R}{3\pi r}$}\,\int_0^\infty \Phi(ix)  \sin (x\frac{r}{R})\,x\,dx\,
\hspace{-36mm}
\hspace{-16mm}&& \hbox{\Large $\big\Updownarrow$}\ \ \ \hbox{duality}\ {\cal D}
\vspace{3mm}\\
&\hspace{-10mm} \searrow\!\!\!\!\!\!\nwarrow\  \hbox{(Fourier)}
\vspace{0mm}
\\
&&\hspace{-14mm}
\underline{\hbox{\it form factor}}: 
\dis\hspace{16mm} f(k)\,=\, \Phi(ix)\,=\, \langle\,e^{i\vec k.\vec r}\,\rangle\ =\
\langle \, \frac{\sin kr}{kr}\, \rangle\ =  \, \dis \sum_0^\infty\ 
\frac{(-1)^n\ \langle \,r^{2n}\rangle}{(2n+1)! \ R^{2n}}\ \,x^{2n}
\ea
$}$
\vspace{2mm}\\ 
\ea 
$
} 
\end {table}

For an homogeneous sphere,  with
$
\langle \, r^{2n}\, \rangle =  \hbox{\small $\dis \frac{3}{2n+3}$}\, R^{2n}\,$ i.e.
\be
\frac{\langle \, r^{2n}\, \rangle}{R^{2n}} =  A_{2n}= \frac{3}{2n+3} = 1, \ \,\frac35,\ \,\frac37,\ \,\frac13, \ ...\ , 
\ee
we get
\be
 \label{hol1}
 \left\{\ \ba{ccccccc}
 \phi(x)\!&=&\!\dis \langle \, \frac{\sinh kr}{kr}\, \rangle  &=&\dis \sum_0^\infty\ \frac{x^{2n}}{(2n+1)!}\ \frac{3}{2n+3} &=&\dis 
 1\, +\, \frac{x^2}{10}\, +\,\frac{x^4}{280}\,    +\,\frac{x^6}{15\,120}   +\,...\  ,
 \vspace{2mm}\\
 \tilde \phi(x)\!&=&\!\dis  \langle \, \frac{\sin kr}{kr}\, \rangle  &=&\dis \sum_0^\infty\ (-1)^n\ \frac{x^{2n}}{(2n+1)!}\ \frac{3}{2n+3} &=&\dis 
 1\, -\, \frac{x^2}{10}\, +\,\frac{x^4}{280}\,    -\,\frac{x^6}{15\,120} +\,...\  .
 \ea\right.
 \ee
This is easily identified with the expressions (\ref{hom}) of the hyperbolic and ordinary form factors of an homogeneous sphere,
\be
\left\{\  \ba{ccccc}
\phi(x)\!&=&\! \dis \frac{3\,(x\,\cosh x-\sinh x)}{x^3} &=&  \hbox{\small $\dis 3\ \sum_0^\infty \ \frac{x^{2n}}{(2n+2)!}\ \left[\,1-\frac{1}{2n+3}\,\right]$}\ ,
\vspace{1mm}\\
\tilde\phi(x)\!&=&\! \dis \frac{3\,(\sin x - x\cos x)}{x^3}&=& \hbox{\small $\dis 3\ \sum_0^\infty \ (-1)^n\ \frac{x^{2n}}{(2n+2)!}\ \left[\,1-\frac{1}{2n+3}\,\right]$}\ .
\ea \right.
\ee

\vspace{2mm}

The form factor of an homogeneous sphere
van\-ishes and changes sign for $\,\tan x = x$, in connection with a spherically symmetric solution for the inside potential ${\cal V}_{\rm in} (r)$ for which  ${\cal V}_{\rm out}(r) $ vanishes identically (ensuring a resonant behavior in the presence of a time-dependent external excitation). The inside potential (obeying the second equation (\ref{poiss2})) and field are then obtained as
\vspace{-2mm}

\be
\label{potin}
{\cal V}_{\rm in}(r)\,= \,\frac{\rho_0}{k^2}\, 
\left[ \  \hbox{\Large $\frac{\sin kr / kr}{\sin kR/kR}$} -1\ \right]\,,
\ \ \ 
{\cal E}_{\rm in}(r)=-\frac{d\ }{dr }\,{\cal V}_{\rm in}(r) = \frac{\rho_0}{k^2 r}\,\frac{1}{\sin kR / kR} 
\,\left[\,\frac{\sin kr}{kr} -\,\cos kr\,\right]\ .
\ee
The field vanishes at $r=0$, and behaves at small $r$, in the small $k$ limit,  like ${\cal E}_{\rm in}(r) \simeq \rho_0 \,\frac{r}{3} = [\rho_0\,\frac{4\pi r^3}{3}]/$ $(4\pi r^2)$, in agreement with Gauss theorem which applies in this limit. Its vanishing 
at $r\!=\!R\,$, allowing for a vanishing external potential,  
requires $\tan x=x\,$, i.e. $x=kR\simeq 
4.49,\  7.73,\ 10.90$,  ...\ . 
\,These zeroes of $\phi(x)$ ultimately approach $x\simeq (n+\frac12)\,\pi$, corresponding to $\cos kR \simeq 0$ and $(n+\frac12)\,\lambda' \simeq 2R\,$ with a wave- length $\lambda' =2\pi/k = 2\pi \lambda $.

\vspace{2mm}

The relations between $\rho(r)$ and its moments 
with the hyperbolic and ordinary form factors 
are represented in Table \ref{relations}.
For the Earth, with a moment of inertia $I = \frac23\, M \,\langle\, r^2\,\rangle  \simeq .3308\,MR^2\,$, 
the correspondence 
\vspace{1mm}
between form factors reads
\be
\Phi(x)= \langle\frac{\sinh kr}{kr}\rangle \simeq \,1+ .0827\,x^2 + .0027\,x^4 +\, ... \ \ \ \hbox{\large$\stackrel{\cal D}{\longleftrightarrow}$}\ \ \ \tilde\Phi(x)=  \langle\frac{\sin kr}{kr}\rangle \simeq\, 1-.0827\, x^2+ .0027\,x^4 +\, ...\, .
\ee

\section{\boldmath \ The effective density $\bar\rho(x)$ as a decreasing function of $x=R/\lambda$}
\label{sec:dec}

For each range $\lambda$ corresponding to $x= kR = R/\lambda$  we define an effective density
$\bar\rho(x)$,  such that the Earth generates, for this range $\lambda$,
the same Yukawa potential as an homogenous sphere with density $\bar\rho(x)$. It is expressed as 
\vspace{-7mm}

\be
\label{rhobar0}
\bar\rho(x) \,=\,\rho_0 \ \frac{\Phi(x)}{\phi(x)} \,=\, 
\frac{\hbox{\small $\displaystyle\int_0^R$}\ \rho(r)\ rdr \, \sinh kr}
{\hbox{\small $\displaystyle\int_0^R$}\ rdr \, \sinh kr}\ .
\ee
When the range $\lambda = 1/k$ decreases, this effective density gets more and more sensitive to the densities $\rho(r)$ within the external shells of the sphere. It decreases, for $\rho(r)$ decreasing with $r$, down to the density $\rho(R)$ within the most external shells.
The decreasing of $\bar\rho(x)$ for increasing $x$, when $d\rho(r)/dr<0$, may be verified by deriving  $\bar\rho(x)$ with respect to $k= 1/\lambda$, evaluating
\be
\frac{d\bar \rho}{dk}\ =\ \frac{\hbox{$ \int_0^R \!\!\int_0^R$} \, \rho(r)\ rdr\ r'dr'\ (r\cosh kr \sinh kr'-r'\sinh kr\cosh kr')}{\left[\,\hbox{$\int_0^R$}\ rdr \, \sinh kr\,\right]^2}\ =\ \frac{N}{D}\ .
\ee
The numerator reads
\be
N\,=\,\hbox{\small$\displaystyle \int_0^R\!\!\!\!\int_0^{R}$} \ [\rho(r)-\rho(r')]\ rdr\ r'dr'\ (r\cosh kr\, \sinh kr')\ ,
\ee
\vspace{-4mm}

\noindent
with
\be
\ba{c}
r\,\cosh kr \sinh kr'\,=\,\frac{1}{4}\ [(r+r') + (r-r')]\ [\,\sinh k(r+r') - \sinh k(r-r')\,]
\vspace{5mm}\\
= 
\frac{1}{4}\, (r^2-r'^2)\, \left[ \hbox{\Large $ \,\frac{\sinh k(r+r')}{r+r'} $} -  \hbox{\Large $\frac{\sinh k(r-r')}{(r-r')}$}\right]+\hbox{symmetric terms},\!\!
\ea
\ee
\vspace{-2mm}

\noindent
so that
\be
N\,=\,\hbox{$\displaystyle  \int_0^R\!\!\!\!\int_0^{R}$} \, [\,\rho(r)-\rho(r')\,]\ (r^2-r'^2)\ \displaystyle \left[\,\frac{\sinh k(r+r')}{r+r'}  - \frac{\sinh k(r-r')}{(r-r')}\,\right]
 \frac{rdr\ r'dr'}{4}\ .
\ee
For large $\lambda$ i.e. small $m=k$, $ \bar\rho \to\rho_0$ and $d\bar\rho/dk\to 0$, in agreement with eq.\,(\ref{rhobar0}). For a density $\rho(r)$ decreasing with $r$ the product of the first two factors in $N$  is negative while the third one within the brackets is positive, so that $N <0\,,\ d\bar\rho/dk <0\,$ i.e. $d\bar\rho/dx <0$\, (or $d\bar\rho/d\lambda > 0$). $\bar\rho(x)$ decreases when the range $\lambda$ gets smaller, as expected since  the potential gets preferentially induced by the more external shells of the sphere, with a lower density.

\vspace{2mm}
For larger $x$ i.e. smaller $\lambda$'s the regions at a depth $R\!-\!r \approx \lambda$ below surface are essential  to  generate the Yukawa potential, so that $\bar\rho(x)$ is  largely determined by the density at depth $\lambda$. This leads, for sufficiently regular expressions of the density, to expect $\bar\rho(x)\approx \rho(R-\lambda)$. And we get, ultimately,
\be
\bar\rho(x)\ \ \stackrel{\stackrel{\hbox{\footnotesize large $x$}}{}}{\longrightarrow} \ \ \rho(R)\ .
\ee

\section{Form factors for several density distributions}
\label{sec:hyp}

\subsection{\ The fundamental of an hydrogen atom}
\label{hydro}

Let us consider an exponentially decreasing density distribution $\rho(r)$ proportional to $e^{-r/R}$, written as $e^{-2r/a}/(\pi a^3)$ with $R=a/2$ as for the fundamental of an hydrogen atom, with wave function $e^{-r/a}/(\pi a^3)^{1/2}$ where $a \simeq .529\, \times 10^{-10}$\,m\,  is the Bohr radius. 
The form factors are evaluated,  assuming $k < 2/a$ (i.e. $x=kR <1$) in the hyperbolic case so that $\rho(r)$ decreases faster than $e^{-kr}$,
allowing for $\,g(k) = \langle\, \frac{\sinh kr}{kr}\, \rangle$ to be finite. We have
\be
g(k) = \frac{\int_0^\infty\, [\,e^{-(\frac2a-k)\,r} - e^{-(\frac2a+k)\,r}]\, rdr}{2k\,\int_0^\infty e^{-2r/a}\  r^2dr}
=\frac{(2/a)^2}{2ka} \left[ \frac{1}{\left(\frac2a\!-\!k\right)^2} -  \frac{1}{\left(\frac2a \!+\!k\right)^2}  \right]
=\frac{1}{\left(\,1-\frac{k^2a^2}{4}\,\right)^2}= \frac{1}{\left(\,1-x^2\,\right)^2}\, ,
\ee
and, changing $k^2\to -\,k^2$ or $x^2\to -\,x^2$ (now for all $k$ or $x$), 
\be
\label{h3}
f(k) = \langle\, \frac{\sin kr}{kr}\, \rangle =\,\frac{1}{\left(\,1+\frac{k^2a^2}{4}\,\right)^2}= \,\,\frac{1}{\left(\,1+x^2\,\right)^2}\ .
\ee
This is the form factor
for an electron in the fundamental state of an hydrogen atom with a momentum transfer $k = (2\pi/\lambda' ) \times 2\,\sin\theta/2\,$.

\vspace{2mm}
These form factors may be recovered from the moments
\be
\label{h1}
\langle\ r^{2n}\ \rangle \,=\,\frac{\int_0^\infty e^{-2r/a}\ r^{2n}\, r^2dr}{\int_0^\infty e^{-2r/a}\  r^2dr}\,=\,a^{2n}\ \frac{(2n+2)!}{2^{(2n+1)}}\ ,
\ee
including $\langle\, 1/r \,\rangle \!= 1/a $, $\langle\, r^2 \,\rangle\! = 3\,a^2$\,, etc.. This  provides the power series expansions \footnote{This corresponds to 
$\, \sum_0^\infty u^n ={1}/(1-u) $ and, 
\vspace{-.1mm}
by derivation, $\, \sum_0^\infty (n\!+\!1)\,u^n =  1+2u+3u^2 + ... = {1}/{(1-u)^2} $, 
still conven\-tionally taken as as valid for $u\neq 1$, even when the series is no longer absolutely convergent.}, absolutely convergent for $x<1$, and resummed into
\vspace{-3mm}

\be
\label{h2}
g(k) = \,\langle\, \frac{\sinh kr}{kr}\, \rangle 
=\,\sum_0^\infty\, \frac{\langle\, (kr)^{2n}\,\rangle}{(2n+1)!}\,
=\,
\sum_0^\infty\, \left(\frac{k^2a^2}{4}\right)^{\!n}\! (n+1)\,=\,\frac{1}{\left(\,1-\frac{k^2a^2}{4}\,\right)^2}= \, \frac{1}{\left(\,1-x^2\,\right)^2}\ .
\ee
And, changing $k^2\to -\,k^2$ or $x^2\to -\,x^2$,
\be
\label{h3bis}
f(k) = \langle\, \frac{\sin kr}{kr}\, \rangle 
\,=\,
\sum_0^\infty\ (-1)^n\ \left(\frac{k^2a^2}{4}\right)^{\!n} (n+1)\,=\,\frac{1}{\left(\,1+\frac{k^2a^2}{4}\,\right)^2}= \, \frac{1}{\left(\,1+x^2\,\right)^2}\ .
\ee

\subsection{A gaussian density distribution}

\vspace{-1mm}

In a similar way we evaluate, for a gaussian density distribution
$
\rho(r) = \hbox{\small$\dis \frac{1}{(\sigma \sqrt{2\pi})^3}$}\ e \hbox{\large$\dis ^{-\frac{r^2}{2\sigma^2}}$}\,,
$
the moments
\vspace{-1mm}

\be
 \langle\,r^{2n}\,\rangle 
 \ = \ [\,1\times 3 \times ...\times (2n\!+\!1)\,] \ \sigma^{2n}=\ \frac{(2n+1)\,!}{2^n\,n\,!}\
\sigma^{2n}\,,
\ee
so that we recover, from the power series expansion of the hyperbolic form factor,
\be
g(k) \,=\,\langle\ \frac{\sinh kr}{kr}\ \rangle\,=\,\sum_0^\infty\ \frac{k^{2n}\, \langle\,r^{2n}\,\rangle}{(2n+1)\,!}
\,= \,\sum_0^\infty \ \frac{1}{n\,!} \, \left(\frac{\sigma^2 k^2}{2}\right)^{\! n}\,= \,\dis e^{\,\hbox{$\frac{\sigma^2 k^2}{2}$}}\ .
\ee
\vspace{-2mm}

\noindent
We then get again, by duality, the form factor
$
f(k) = g(ik) = e^{\,\hbox{$-\frac{\sigma^2 k^2}{2}$}}
$,
which is the Fourier transform of $\rho(r)$\,.

\subsection{\boldmath A density proportional to $\,1/r$}

Let us consider a density distribution proportional to $1/r$ for $r<R$ with average density $\rho_0$, namely
\be
\rho(r) = \rho_0\ \frac{2R}{3\,r}\ \ \hbox{ for} \ r<R\,.
\ee
It is integrable to unity, and decreases  with $r$, down to $\frac23\ \rho_0$ near the surface of the sphere.
The hyperbolic form factor is given by
\be
\Phi(x)= \langle\ \frac{\sinh kr}{kr}\ \rangle \,=\, \frac{2}{R^2}\int_0^R \,\frac{\sinh kr}{kr} \ \frac{1}{r}\ r^2 dr \,=\,\frac{2}{x^2}\, \int_0^x \sinh u \ du \ ,
\ee
i.e.
\vspace{-4mm}
\be
\label{1/r0}
\framebox [7.2cm]{\rule[-.35cm]{0cm}{.95cm} $ \dis
\Phi(x) \,=\, \frac{2}{x^2}\ (\cosh x -1) \,=\, \left(\frac{\sinh x/2}{x/2}\right)^2\,.
$}
\ee
\vspace{2mm}

The form factor for this density distribution is obtained by changing $x^2 \to -\,x^2$, so that
\be
\label{1/r}
\tilde \Phi(x)= \Phi(ix) = \langle\ \frac{\sin kr}{kr}  \ \rangle =\, \frac{2}{x^2}\ (1-\cos x) = \left(\frac{\sin x/2}{x/2}\right)^2\,.
\ee
It is $\,\geq 0$, and vanishes for $x= 2n\pi$,
in connection with an inside solution for the potential for which the outside potential vanishes,
\vspace{-6mm}

\be
{\cal V_{\rm in}}(r)=\rho_0\ \frac{2R}{3}\  \frac{\cos kr-1}{k^2r}\,.
\ee
This potential is $\leq 0$,
\vspace{-1.2mm}
and verifies the Poisson equation in (\ref{poiss2}), $(-\triangle-k^2)\, {\cal V}_{\rm in}(r) = \hbox{\small$\dis \frac{1}{r}$} \,(\hbox{\footnotesize$\dis -\frac{\partial^2\ }{\partial r^2}$}\!-k^2) \,r\,{\cal V}_{\rm in}(r)= \rho_0\,\frac{2R}{3r}= \rho(r)$.
\vspace{-.6mm}
 It vanishes at the origin, and, together with its derivative, at $r=R$, for $x=$ $ kR =2n\pi$, i.e.  $R = n\, \times\,$(wavelength $\lambda'= 2\pi /k$). This allows for a vanishing external potential ${\cal V_{\rm out}}(r)$, with $\tilde \Phi(x=kR=2n\pi)$ $=0$. \, 
The corresponding field
\be
{\cal E}_{\rm in}(r)=\,-\,\frac{d\ }{dr\,}\,{\cal V_{\rm in}}(r) = \,\rho_0\,\frac{2R}{3}\left(\,\frac{\sin kr}{kr}+\frac{\cos kr-1}{k^2r^2}\,\right)\,,
\ee
is still singular at the origin, 
\vspace {-.4mm}
with a non-zero value
${\cal E}_0\!= \rho_0\, \frac{R}{3}= [\,\rho_0\, \frac{4\pi r^2 R}{3}\,]/ \,(4\pi r^2)$. 
This one is equal to the charge $Q_r\!=\! \rho_0\, 4\pi r^2 R/3$ within the sphere of radius $r$, divided by its surface, in agree\-ment with Gauss theorem which applies in the small $r$ and small $k$ limit. 
\vspace {-.3mm}
This corresponds, as in eq.\,(\ref{potin}) for a constant density, to  ${\cal V}_{\rm out} (r)=0$, i.e. to a stationary wave with angular frequency $\omega = \pm\, k$, 
\vspace {-.3mm}
proportional to $\cos\, (kt +\varphi) $, confined within the sphere.
 The zeros of $\tilde \Phi(x)$ (or poles of $1/\tilde\Phi(x)$) correspond 
 to a resonant behavior of the inside solutions in the presence of an external potential oscillating with time.

\vspace{2mm}

The moments of the density distribution are
\be
\langle\,r^{2n}\,\rangle \,=\,\ \frac{\int_0^R \,r^{2n+1} dr}{\int_0^R\, rdr} \,=\, \frac{R^{2n}}{n+1}\ ,
\ee
including
$
\langle\,r^{2}\,\rangle = {R^2}/{2}\,,\  \langle\,r^{4}\,\rangle = {R^4}/{3}\,,\  \langle\,r^{6}\,\rangle = {R^6}/{4}\,,\  ...\,
$.
 This provides the expansion of $\Phi(x)$ as
\be
\ba{ccc}
\Phi(x)&\simeq &\dis \sum_0^\infty\ \frac{x^{2n}}{(2n+1)!}\  \frac{1}{n+1} \,=\, \sum_0^\infty\  \frac{2}{(2n+2)\,!}\ x^{2n}\, = \,1 + \frac{x^2}{12}+ \frac{x^4}{360}+ \frac{x^6}{20\,160}\,+ \frac{x^8}{1\,814\,400}\,+\,\,...
 \vspace{2mm}\\
&\simeq &
\dis 
\,1 \,+ \,.0833\ x^2 + 2.78\times 10^{-3}\ x^4 + 4.96 \times 10^{-5}\ x^6 + 5.51 \times 10^{-7}\,+\,\,...\,.
\ea
\ee
The $\,{x^2}/{12}\,$ term corresponds to a moment of inertia $I = \frac23 \,M\langle\,r^{2}\,\rangle = \frac13\, MR^2$.
$\Phi(x)$  is naturally smaller than the $\phi(x)$  for an homogeneous sphere, given in eq.\,(\ref{hol1}).
This 
leads to the effective density
\be
\framebox [7.3cm]{\rule[-.4cm]{0cm}{1.05cm} $ \dis
\bar\rho(x) = \rho_0\ \frac{\Phi(x)}{\phi(x)} = \rho_0\ \frac{2x\,(\cosh x -1)}{3\,(x\cosh x-\sinh x)}\ \,,
$}
\ee
decreasing from $\bar \rho(0) =\rho_0$ for a long range force, down to $\rho(R) =  \frac{2}{3}\ \rho_0$, as represented later, in Fig.\,\ref{rhorho} of Sec.\,\ref{sec:earth}, in dotted red.

\subsection{\boldmath Densities proportional to $r^p$}

Let us now consider  a density $\rho_p(r)$ proportional to $r^p$ for $r\leq R$ (with $p > -3$ for integrability  at the origin), i.e.
\vspace{-6mm}

\be
\rho_p(r) = \rho_0\  \, \frac{p+3}{3}\ \frac{r^p}{R^p}\ \ \ \ \hbox{ for} \ r\leq R\,,
\ee
such that $\int_0^R \rho(r) \ 4\pi r^2 dr = \rho_0 \times 4\pi R^3/3$.
It includes the densities
\vspace{-2mm}

\be
\rho_{-2}=\rho_0\  \frac{R^2}{3\,r^2}\,,\ \ \,\rho_{-1}=\rho_0\  \frac23 \frac{R}{r}\,,\ \ \ \,\rho_{0} \,,
\ \ \ \,\rho_{1}= \rho_0\  \frac43 \frac{r}{R}\,,\ \ \,\rho_{2}= \rho_0\  \frac53 \frac{r^2}{R^2}\,,\ \ \ ...\,,
\ee
with the moments 
\vspace{-6mm}

\be
\label{moments}
\langle\,r^{2n}\,\rangle =\ \frac{\int_0^R \,r^{p+2n+2}\  dr}{\int_0^R\, r^{p+2}\  dr} = R^{2n}\ \frac{p+3}{{p+2n+3}}\ ,
\ee
so that, in particular,
\be
I\,=\, \frac23\,M\,\langle \,r^2\,\rangle\, =\, \frac23\ \frac{p+3}{p+5}\ MR^2 \,=\,\left\{\ \frac29\, ,\ \frac13\, ,\ \frac25\, ,\ \frac49\,,\ \frac{10}{21}\,, \,...\ \right\}\, \times MR^2\ .
\ee
Eqs.\,(\ref{ex2},\,\ref{moments}) lead to the expansions of the hyperbolic form factors as
\be
\label{expPhip}
\Phi_p(x)\,=\,\sum_0^\infty\, \frac{x^{2n}}{(2n+1)!}\  \,\frac{p+3}{{p+2n+3}}\, = \,1\, +\, \frac{(p+3)\,x^2}{6\,(p+5)}
\,+ \frac{(p+3)\,x^4}{120\,(p+7)}\, + \frac{(p+3)\,x^6}{5040\,(p+9)}\, +\ ...\ .
\ee
\vspace{2mm}

\vbox{
$\Phi_p(x)$ is directly obtained as 
\vspace{-2mm}

\be
\label{rec0}
\Phi_p(x)\,=\, \langle\ \frac{\sinh kr}{kr}\ \rangle\, =\, \int_0^R \,\frac{\sinh kr}{kr} \ \frac{(p+3)\,r^{p+2}\,dr}{R^{\,p+3}}\,=\ 
\frac{p+3}{x^{p+3}} \ \underbrace{\int_0^x \,\sinh u \ u^{p+1}\,du}_{\hbox{$J_p(x)$}}\ .
\ee
}
\vspace{-5mm}

\noindent
$J_p(x)$ satisfies,  for $p\geq 0$,
the recurrence relation
\vspace{-3mm}

\be
\label{rec}
J_p (x)\,= \, x^{p+1}\, \cosh x - (p+1)\,x^p\, \sinh x +  p\,(p+1)\,J_{p-2}(x)\ .
\ee
\noindent
With $\,J_{-1}(x) =\cosh x -1\,$ it leads,  for $p = 0,\,1,\,2,\,...\,$, \,to $J_0(x)= x\cosh x-\sinh x,\ J_1(x) = x^2\cosh x -2x\sinh x + 2\,(\cosh x -1)$,
 $\,J_2(x) = x^3\cosh x -3x^2\sinh x + 6\, (x\cosh x-\sinh x) $,\  ...\ :
\be
\label{eqPhi0}
\left\{\ 
\ba{ccccl}
J_{-2}(x) &=&\int_0^x\,\hbox{\small$\dis \frac{\sinh u}{u}$}\ du &=&\dis \hbox{Sih}(x) 
 \vspace{1mm}\\
J_{-1}(x) &=&  \int_0^x\,\sinh u\ du  &=&\dis \cosh x -1\ ,
\vspace{2mm}\\ J_{0}(x) &=& \int_0^x\,\sinh u\ u \,du  &=&\dis  x\,\cosh x -\,\sinh x\  ,
\vspace{2mm}\\
J_{1}(x) &=&\int_0^x\,\sinh u\ u^2 du  &=&\dis x^2  \cosh x - 2x\sinh x +2\,(\cosh x-1)\ ,
\vspace{2mm}\\
J_{2}(x) &=& \int_0^x\,\sinh u\ u^3 du  &=&\dis x^3  \cosh x - 3x^2\sinh x +6\,(x\cosh x-\sinh x)\ ,
\vspace{2mm}\\
&&& ...
\ea
\right.
\ee

\noindent
They satisfy
\vspace{-6mm}

\be
\label{eqPhi}
\frac{d \ }{dx} \ \left[\,\frac{x^{p+3}}{p+3}\ \Phi_p(x)\,\right] \,=\, \frac{d \,J_p(x)}{dx} \,= \ \sinh x\ x^{p+1}\ .
\ee

\vspace{1mm}

\noindent
This provides the hyperbolic form factors 
$\Phi_p(x) =  \hbox{\small $\dis  \frac{p+3}{x^{p+3}}$} \ J_p(x)$,   expanded according to eq.\,(\ref{expPhip}) as
\be
\label{list}
\left\{
\ba{cclcl}
\Phi_{-2}(x)\! \! &=&\dis\frac{1}{x}\ \,\hbox{Shi}(x) &=&\dis   1 + \frac{x^2}{18}+ \frac{x^4}{600}+ \frac{x^6}{35\,280}+ ...  \,,\vspace{2mm}\\
\Phi_{-1}(x) \! \! &=&\dis \frac{2}{x^2}\ [\,\cosh x -1\,] &=&\dis 
 1 + \frac{x^2}{12}+ \frac{x^4}{360}+ \frac{x^6}{20\,160}+ ...\,,
\vspace{2mm}\\ 
\Phi_{0}(x)\!\!   &=&\dis  \frac{3}{x^3}\ [\,x\,\cosh x -\,\sinh x\,] \ = \ \phi(x)&=&\dis   1 + \frac{x^2}{10}+ \frac{x^4}{280}+ \frac{x^6}{15\,120}+ ... \,, \vspace{2mm}\\
\Phi_{1}(x) \! \! &=&\dis \frac{4}{x^4}\ [\,x^2  \cosh x - 2x\sinh x +2\,(\cosh x-1)\,] &=&\dis  
1 + \frac{x^2}{9}+ \frac{x^4}{240} + \frac{x^6}{12\,600} + ...\, ,
\vspace{3mm}\\
\Phi_{2}(x) \! \! &=&\dis \frac{5}{x^5} \  [\,x^3 \cosh x - 3x^2\sinh x +6\,(x\cosh x- \sinh x)\,] 
 &=&\dis  
1 + \frac{5 x^2}{42}+ \frac{x^4}{216} + \frac{x^6}{11\,088} + ...\,,\!\!
\vspace{3mm}\\
&& \hspace{40mm}...
\ea
\right.
\ee
where $J_p(x))$ within the brackets verify eq.\,(\ref{eqPhi}). They are expanded as above according to
\be
\left\{\ \ba{ccc}
\Phi_{-1}(x)  &=& \dis \sum_0^\infty\  \frac{x^{2n}}{(2n+1)!}\  \frac{1}{n+1}\, ,
\vspace{1mm}\\
\Phi_{0}(x)  &=&\dis  \sum_0^\infty\  \frac{x^{2n}}{(2n+1)!}\  \frac{3}{2n+3}\, ,
\vspace{1mm}\\
\Phi_{1}(x)  &=&\dis  \sum_0^\infty\  \frac{x^{2n}}{(2n+1)!}\  \frac{2}{n+2}\, ,
\ea \right.
\ \ \ \ 
\left\{\ \ba{ccc}
\Phi_{2}(x)  &=&\dis  \sum_0^\infty\ \frac{x^{2n}}{(2n+1)!}\  \frac{5}{2n+5}\, ,
\vspace{1mm}\\
&& ...
\vspace{1mm}\\
\Phi_p(x)  &=&\dis  \sum_0^\infty\ \frac{x^{2n}}{(2n+1)!}\  \frac{p+3}{p+2n+3}\, ,
\vspace{1mm}\\
&& \ \ \,...\ \ .
\ea \right.
\ee

\vspace{2mm}

We also have
\be
\label{Phi012}
\left\{\ 
\ba{lclclcc}
\Phi_{-1}(x) \!&=&\ \dis \frac{2}{x}\ \ \frac{\cosh x -1}{x}\ ,
\vspace{2mm}\\
\Phi_0(x) = \phi(x)  \!&=&\  \dis \frac{3}{x}\ \frac{d\ }{dx} \, \frac{\sinh x}{x}\!&=&   \dis \frac{3}{x^3}\ (x \cosh x-\sinh x)\ ,
\vspace{2mm}\\
\Phi_1(x) \!&=&\  \dis  \frac{4}{x}\ \frac{d^2\ }{dx^2} \, \frac{\cosh x -1}{x}\!&=& \dis  \frac{4}{x^4}\ [\,x^2  \cosh x - 2x\sinh x +2\,(\cosh x-1)\,]\ ,
\vspace{2mm}\\
\Phi_2(x) \!&=&\  \dis  \frac{5}{x}\ \frac{d^3\ }{dx^3} \, \frac{\sinh x}{x} \!&=&   \dis \frac{5}{x^5}\ 
[\,x^3 \cosh x - 3x^2\sinh x +6x\cosh x-6\sinh x\,] \ .
\vspace{3mm}\\
&&\hspace{20mm}...\\ 
\ea
\right.
\ee
This iterates into
\be
\label{Phi0123}
\left\{ 
\ba{ccccccc}
\hbox{for $p$ even} \ \geq 0: \hspace{-20mm}
\vspace{3mm}\\
\Phi_p(x) \! \!&=&\!\dis \frac{p+3}{x}\ \frac{d^{\,p+1}\ }{dx^{p+1}} \, \frac{\sinh x}{x} \!&=&\!
\dis \frac{p+3}{x^{p+3}}\  [\,x^{p+1} \cosh x- (p+1)\,x^p\sinh x + ... - (p+1)!\,\sinh x\,]\ ,
\vspace{4mm}\\
\hbox{for $p$ odd} \ \geq 1: \hspace{-20mm}
\vspace{3mm}\\
\Phi_{p}(x)  \!\!&=&\! \dis \frac{p+3}{x}\ \frac{d^{\,p+1}\ }{dx^{p+1}} \, \frac{\cosh x -1}{x} \! \!&=&\! \dis \frac{p+3}{x^{p+3}}\ [\,x^{p+1} \cosh x- (p+1) x^p \sinh x+ ... + (p+1)!\,(\cosh x\!-\!1)\, ]\,,
\ea \right.
\ee
in agreement with eqs.\,(\ref{rec0},\,\ref{eqPhi0}-\ref{list}) and with the recurrence relation (\ref{rec}).
At large $x$, $\Phi_p(x)$ behaves as \hbox{$(p+3)\,e^x/ (2x^2)$} and $\phi(x)$ as  $3\,e^x/ (2x^2)$ so that $\bar\rho(x)\to\,\rho_0\ (p+3)/3= \rho_p(R)$, as it should.

\subsection{\boldmath A density vanishing at $r=R$}

To better understand the situation at large $x$ when the density $\rho(R)$ vanishes (for a finite radial extension $R$),  we consider the distribution
\vspace{-4mm}

 \be
 \rho(r)=4\,\rho_0 -3\,\rho_1(r)\, =\, 4\,\rho_0\left(\,1-\frac{r}{R}\,\right)\,,
 \ee
for $r<R$,  with average density $\rho_0$ and decreasing linearly from $4\,\rho_0$ to $0$.
 It leads to
 \be
 \label{Phi01}
 \ba{ccl}
 \Phi(x)&=& \dis 4\,\Phi_0(x) -3\,\Phi_1(x)
 \,=\, \dis  \frac{12}{x^4}\ \left[\,  x\sinh x -2\,(\cosh x-1)\,\right]
 \vspace{2mm}\\
 &=&\dis \sum_0^\infty\  \frac{12\ x^{2n}}{(2n+1)!\,(2n+3)\,(2n+4)} \,=\, 1 + \frac{x^2}{15}+ \frac{x^4}{560}+ \frac{x^6}{37\,800}+ ... \,.
\ea
\ee
The potentially dominant terms at large $x$ i.e. small $\lambda$, that would behave proportionally to $e^x/x^2$,  cancel out as $\rho(R)=0$.
$\Phi(x)$ then behaves like $ 6\, e^x/x^3 = 3\, e^x/2x^2 \times [ \,\rho(R-\lambda)/\rho_0 \!=\! 4 / x\,]$, as if the potential was generated by an homogeneous sphere of density $\rho(R-\lambda$), the density at a depth $\lambda$ below the surface.

\section{\ \boldmath The inversion formula} 
\label{sec:inv}

\subsection{\boldmath Recovering $\rho(r)$ form $\Phi(ix)$}

The hyperbolic form factor $g(k) = \Phi(x=kR)$ in eqs.\,(\ref{ex2},\,\ref{hol0}) encodes the even moments of the density distribution $\rho(r)$. These  are also encoded within the form factor $f(k)$,
\be
\label{dualbis}
f(k) = \tilde \Phi(x)= \Phi(ix)
\,=\,\langle \ \frac{\sin kr}{kr}\  \rangle 
\,=\, \int_0^\infty  \rho(r)\ \frac{\sin kr}{kr}\ 4\pi r^2 dr\,,
\ee
$f(\vec k)$ and $g(\vec k)$ appear as in eqs.\,(\ref{g},\,\ref{d0})  as the 3d Fourier and Laplace transforms of the  density distribution $\rho(r)$, respectively.
They may be reexpressed in spherical coordinates by extending $\rho(r)$ to negative values of $r$ through $\rho(-r) = \rho(r)$, providing
$\rho(r)\!=\! \rho(|r|)$ as an even function of $r$, still normalized in the same way through
\vspace{-4mm}

\be
\hbox{\small $\dis \int_{-\infty}^{\,\infty}$} \rho(r)\ 2\pi r^2 dr= \hbox{\small$\dis \int_0^\infty\rho(r)$}\ 4\pi r^2 dr\,=\,1\ .
\ee
We then have
\be
\label{lap2}
 ik f(k) = 
\,\int_{-\infty}^{\,\infty}\! 2\pi  r\rho(r)\ e^{ikr}\, dr\,=\,ik\,g(ik)\,,
\ \ \ 
k g(k) = 
\,\int_{-\infty}^{\,\infty}\! 2\pi r\rho(r)\ e^{kr}\,dr\,,
\ee
allowing for a smooth treatment of the singularity at the origin.
With $e^{(i)kr}= 1 + (i)  kr + ...\,$ we get back $g(0)= f(0)= \int_0^{\,\infty} \rho(r)\,4\pi r^2dr  
=1$, for a density distribution normalized to unity.

\vspace{2mm}

The odd function $kg(k)$ in eq.\,(\ref{lap2}) appears as the bilateral Laplace transform of the odd function $2\pi r\rho(r)$, just as 
$ikf(k)$ is its Fourier transform. 
One can recover $2\pi r\rho(r)$ through two transformations amounting to an inverse bilateral Laplace transformation, as illustrated 
\vspace*{1.5mm}
by the chain of relations 
\be
\ba{c}
2\pi r\rho(r)\,  \stackrel{\stackrel{\hbox{\small Laplace}}{}}{\longrightarrow}\,
\left(\ba{c}kg(k)  \\ =  \\ k\,\Phi(x\!=\!kR) \ea\right) \,
\ba{c}
\hbox{\small Duality}
\vspace{-1mm}\\
\longleftrightarrow
\vspace{-1mm}\\
\cal D
\ea
\,
\left(\ba{c}ikf(k)\\ = \\ ik\ \tilde\Phi(x\!=\!kR) \ea\right)
\ \stackrel{\stackrel{\hbox{\small Fourier}}{\ }}{\longleftrightarrow}\ \,2\pi r\rho(r)\,.\,
\vspace{4.5mm}\\
g(k)= \hbox{\em hyperbolic form factor}\hspace{10mm}   f(k) = \hbox{\em form factor} \hspace{14mm}
\vspace{1mm}\\
= f(-ik)  \hspace{38mm} = g(ik) \hspace{9mm}
\ea
\ee
The bilateral Laplace transform $ {\cal L}$ acts on $2\pi r \rho(r)$ considered as an odd function of $r$, turning it into $kg(k)$. This one is changed into $ikf(k)$ through a duality transformation ${\cal D}$ 
(squaring to unity), under which $k \leftrightarrow i\,k$. This $ikf(k)$ is then identified as the direct Fourier transform of $2\pi r \rho(r)$. 
More generally the bilateral Laplace transform $ {\cal L}$ of a density distribution $\rho(\vec r)$, and its inverse $ {\cal L}^{-1}$, can be expressed in terms of the Fourier transform ${\cal F}$ and its inverse ${\cal F}^{-1}$, using the duality transformation ${\cal D}$, 
according to 
\be
{\cal L} \,=\,  {\cal D}\ {\cal F}\,,
\ \ \ 
{\cal L}^{-1} ={\cal F}^{-1}\,{\cal D}\ ,
\ee
with
\vspace{-3mm}
\be
{\cal F}^{-1}\,{\cal D}\  {\cal L}\, \,= 1\!\! 1\ .
\ee

\vspace{0mm}

In spherical coordinates we recover $2\pi r\rho(r)$ from the inverse Fourier transform of $kf(k)$, so that
\be
\label{ext}
r\rho(r)\,=\, \frac{1}{4\pi^2} \int_{-\infty}^\infty ikf(k) \ e^{-ikr}\,dr\,=\, \frac{1}{4\pi^2}\int_{-\infty}^\infty kf(k) \, \sin kr\,dr
\ee
or directly in terms of the analytic continuation of 
 $\Phi(ix)= g(ik)= f(k)$, as
\be
\label{inv}
\rho(r)\,=\, \frac{1}{2\pi^2 r}\,  \int_0^\infty kg(ik) \, \sin kr\,dk \,=\,\frac{1}{2\pi^2R^2 r}\, \int_0^\infty \Phi(ix) \, \sin\hbox{\large$\left(\right.$}x\frac{r}{R}\hbox{\large$\left.\right)$}\ x\,dx\,.
\ee
This may be rewritten in terms of the average density  $\rho_0$ on the sphere of radius $R$, with
\be
\label{invbis}
\framebox [7.2cm]{\rule[-.35cm]{0cm}{.95cm} $ \dis
\rho(r)\,=\, \rho_0 \ \,\frac{2R}{3 \pi r}\, \int_0^\infty
 \Phi(ix) \, \sin\hbox{\large$\left(\right.$}x\frac{r}{R}\hbox{\large$\left.\right)$}\ x\,dx\,.
$}
\ee

\subsection{\boldmath Verification, with interesting examples}

\vspace{-1mm}

As a check 
\vspace{-.7mm}
we can insert in eq.\,(\ref{invbis}) the expression of $\,\Phi(ix)= $ \hbox{\Large $\langle\,\frac{\sin kr}{kr}\,\rangle $} =  \hbox{\large $\int_0^R $} \hbox{\Large $ \frac{\rho(r')}{\rho_0}\, \frac{\sin kr'}{kr'}\, \frac{3r'^2dr'}{R^3}$},
which provides back $\rho(r)$, recovered as 
\be
\label{inv2}
\ba{ccl}
\rho(r) \!&\stackrel{\hbox{?}}{=} &\!\dis
\rho_0\ \frac{2R}{3\pi r}\,\int_0^R \frac{\rho(r')}{\rho_0} \ \frac{3r'^2dr'}{R^3}\,\int_0^\infty\,
\frac{\sin\hbox{\large$\left(\right.$}x\,\frac{r'}{R}\hbox{\large$\left.\right)$}}{x\,\frac{r'}{R}}\  \sin\hbox{\large$\left(\right.$}x\,\frac{r}{R}\hbox{\large$\left.\right)$}\ x\,dx\ 
\vspace{2mm}\\
\!&=&\!
\hbox{\small $\dis \frac{R}{\pi r}\,\int_0^R \rho(r') \ \frac{r'dr'}{R^2}\,
\underbrace{\int_0^\infty
\left(\,\cos\hbox{\large$\left(\right.$}x\,\frac{r-r'}{R}\hbox{\large$\left.\right)$}-\cos\hbox{\large$\left(\right.$}x\,\frac{r+r'}{R}\hbox{\large$\left.\right)$} \,\right)\ dx}_{\hbox{\normalsize $ \pi\, [\,\delta (\frac{r-r'}{R})- \delta (\frac{r+r'}{R})\,]$} } $}
\,=\,\rho(r)\ .
\ea
\ee

\vspace{1mm}

For example with $\rho(r)=\frac23\, \rho_0 \, R/r$ for $r<R$, 
we have as in eq.\,(\ref{1/r0}) $\,\Phi(x) =\frac{2}{x^2}\,(\cosh x-1)$
and 
$\Phi(ix) =$ $\frac{2}{x^2}\,(1-\cos x)$, so that the inversion formula (\ref{invbis}), that we intend to verify, reads
\vspace{-.5mm}
\be
\ba{ccc}
\rho(r) \!&\stackrel{\hbox{?}}{=} & \! \dis\rho_0\ \frac{2R}{3\pi r}\, \int_0^\infty  \frac{2}{x^2}\, (1-\cos x)\ \sin\, (x\,\frac{r}{R})\ x\,dx
\,=\,
\rho_0\ \frac{4R}{3\pi r}
\,\int_0^\infty\, \frac{1-\cos x}{x}\ \sin\, (x\,\frac{r}{R})\ dx 
\vspace{3mm}\\
 &=& \ \  \dis \rho_0\ \frac{4R}{3\pi r}\,\left[\ \int_0^\infty \frac{\sin x}{x}\ dx  -\int_0^\infty {\frac{\sin \,x\,(1+\frac{r}{R})- \sin\, x\,(1-\frac{r}{R})}{2x}}\ dx\ \right] \ .
\ea 
\ee
The expression within the bracket is equal to \hbox{$\,\int_0^\infty$ \large $\frac{\sin u}{u}$} $du = \pi/2\,$ for $r<R$ and vanishes for $r>R$. This provides back the initial density distribution
\be
\label{1/r2}
\rho(r) \,= \ \left\{\ \ba{ccc}  \dis \rho_0\ \frac{2R}{3r} &\ \ & \hbox{if}\ \ r<R\,,
\vspace{1mm}\\
0 \ \ &\ \ & \hbox{if}\ \ r>R\,.
\ea\right.
\ee

\vspace{2.5mm}

Eqs.\,(\ref{ext},\,\ref{invbis}) may still be used for a pointlike distribution, 
\vspace{.2mm}
provided they are understood at the distribution level. 
\vspace{-.5mm}
\,For a pointlike distribution normalized to unity $\rho(\vec r) = \delta^3(\vec r)$, 
$\Phi(x)
= 1$,
and eq.\,(\ref{ext}) (or equivalently, eq.(\ref{invbis}) with an arbitrary radius $R$, 
and $\rho_0=  1/\frac{4\pi R^3}{3}$ for a distribution normalized to unity) reads 
\vspace{-6mm}
 
\be
\label{inv1}
\rho(r)\, =\, 
\frac{1}{2\pi^2r}\,  \int_0^\infty k \ \sin kr\ dk\,=\, -\,\frac{1}{2\pi r}\ \delta '(r)\ .
\ee 
Indeed  $\,\int_0^\infty k \, \sin kr\, dk =$ 
$ -\ \frac{d\ }{dr} \int_0^\infty \, \cos kr\, dk = -\,\frac12\, \frac{d\ }{dr} \int_{-\infty}^{\ \infty} \, e^{ikr}\, dk =\!  -\,\pi\ \delta'(r)$, 
\vspace{-.4mm}
which leads to the above non-conventional expression of $\rho(r)$ in eq.\,(\ref{inv1}).

\vspace{2mm}

We thus let the $3d$ space density $\rho(\vec r) = \rho(r) = -\,\frac{1}{2\pi r}\ \delta '(r)$ act on a spherically symmetric test function $f(\vec r)= f(r)$ with compact support and regular at the origin. $f(r)$ is extended to an even function of $r$ continuous and derivable at the origin (with $f'(0)=0$), so that
\be
\langle\, f\,  | \, \rho\, \rangle = 
\int_0^\infty f(r)\  \hbox{\large $\left[ \right. $}-\,\frac{1}{2\pi r}\, \delta'(r)\hbox{\large $\left.\right]$}\ 4\pi r^2 dr = 
\int_{-\infty}^\infty  f(r)\  \hbox{\large $\left[ \right. $}-\,\frac{1}{2\pi r}\, \delta'(r)\hbox{\large $\left.\right]$}\ 2\pi r^2 dr
\,= \dis \left[\,\hbox{\small $ \dis \frac{d\ }{dr}$} \,[rf(r)]\,\right]_{\,r=0} = f(0)\ .\ 
\ee
This provides back
$\rho(\vec r) =  \delta^3(\vec r)
$
as a pointlike distribution at the origin.

\vspace{2mm}

Eq.\,(\ref{inv1}) may be reexpressed as 
\be
\label{delta}
\rho(r) = -\,\frac{1}{2\pi r}\ \delta '(r)\, =\, \frac{1}{2\pi r^2}\ \delta(r) \, -\frac{1}{2\pi r^2}\,  \frac{d\ }{dr}\, [\,r\,\delta(r)\,] \ .
\ee 
The last term does not contribute to $\,\langle \,f\,| \,\rho\, \rangle =\,\int_0^\infty f(r)\,\rho(r)\, 4\pi r^2dr\,$, 
\vspace{-.6mm}
as 
$\,-\,2\, \int_0^\infty f(r)\,\frac{d\ }{dr}\, [\,r\,\delta(r)\,] \ dr\,= $ $–\,\int_{-\infty}^{\, \infty} f(r)\,\frac{d\ }{dr}\,[\,r\,\delta(r)\,]\,dr
=\,\int_{-\infty}^{\, \infty} \frac{df(r)}{dr}\,  r\ \delta(r)\,dr = 0\,$.
This leaves us with 
\be 
\rho(r) \,=\, \frac{1}{2\pi r^2}\ \delta(r) \ .
\ee
This $\rho(r)$, which integrates to
$\int_0^{\infty} \rho(r)\, 4\pi r^2 dr = 2\, \int_0^{\infty} \delta(r) \,dr= \int_{-\infty}^{\, \infty}\, \delta(r) \,dr= 1$ 
\vspace{-.6mm}
and leads to
$\,\langle \,f\,| \,\rho\, \rangle = 2\, \int_0^{\infty} f(r)\, \delta(r) \,dr= f(0)\,$, does correspond to a pointlike distribution at the origin\,\,\footnote{This extends to $d$ space dimensions 
\vspace{-.1mm}
as $\rho(r)=2\  \delta(r)/(S_{d\!-\!1}R^{d\!-\!1})$  where $S_{d\!-\!1}$ is the surface of a $d\!-\!1$-sphere of radius 1 in $d$ dimensions, with $\rho(r)=\frac{1}{\pi r}\, \delta(r)$ in $2d$, $\frac{1}{2\pi r^2}\, \delta(r)$ in $3d$,  $\frac{1}{\pi^2 r^3}\, \delta(r)$ in $4d$,\ ...\,.}.

\section{Simple approximations for the Earth density distribution}
\label{sec:earth}

\subsection{\boldmath A density decreasing linearly from $\,5\,\rho_0/2\,$ to $\rho_0/2$}

Let us now consider more specifically the Earth.
A simple approximation for its density is to consider that it decreases linearly from $\rho_c = \frac52\,\rho_0\simeq 13.775 $ g/cm$^3$ at its center, down to $\rho_e= \frac12\,\rho_0\simeq 2.755 $ g/cm$^3$ near the surface, with
\vspace{-6mm}

\be
\label{rhol0}
\rho_l(r) =  \rho_0 \left(\frac52 -  \frac{2r}{R}\right) \,=\,\frac52\ \rho_0 -\,\frac32 \ \rho_1(r)\ ,
\ee
and average density $\rho_0$. 
The corresponding $\Phi_l(x)$ is, according to eqs.\,(\ref{Phi012}),
\be
\Phi_l(x) \,=\, \frac52\ \Phi_0(x)\,-\frac{3}{2}\ \Phi_1(x)
\,=\, \frac52\ \,\frac{3\,(x\cosh x -\,\sinh x)}{x^3}\,-\,6\ \,\frac{(x^2+2)\cosh x-2x\,\sinh x -2}{x^4}\ ,
\ee
i.e.
\vspace{-6mm}

\be
\label{phil}
\Phi_l(x) \,=\, \dis \frac{3}{2x^4}
\left[(x^2\!-8)\cosh x +3\,x\sinh x+8\,\right]\ .
\ee
It behaves at large $x$ like $3\,e^x/4x^2$, as if it  were generated by an homogeneous sphere of density $\rho_0/2$.

\pagebreak

\vspace{2mm}

Its moments are 
\be
\langle \,r^{2n}\,\rangle  = R^{2n} \ \frac{\int_0^1  (5-4u)\,u^{(2n+2)} du}{\int_0^1 (5-4u)\,u^2 du}\,=\,\frac{\frac{5}{2n+3}-\frac{4}{2n+4}}{\frac23}\,R^{2n}\,=\,\frac{3\,(n+4)}{(2n+3)\,(2n+4)}\,R^{2n}\,,
\ee
including
$
\langle r^{2}\rangle  = \frac12\,R^{2}$,
corresponding to a moment of inertia
$I \!=\!  \frac13 \,MR^2$
(slightly larger than $\,.3308\,MR^2$).
This leads to the expansion
\be
\label{expphil}
\Phi_l(x) \,=\,  \sum_0^\infty\ \frac{3\,(n+4) \ x^{2n}}{(2n+1)!\,(2n+3)\,(2n+4)} \,=\, 1 + \frac{x^2}{12}+ \frac{3\,x^4}{1120}+ \frac{x^6}{21\,600}+ ... \ ,
\ee
average between the expansions of $\Phi_0(x)=\phi(x)$ and $4\,\Phi_0(x)- 3\,\Phi_1(x)$ in eqs.\,(\ref{list},\,\ref{Phi01}).

\vspace{2mm}

Expression (\ref{phil}) of $\Phi_l(x)$ may also be directly expanded as
\be
\label{exprhol}
\ba{ccl}
\Phi_l(x) &=&
\dis \frac{3}{2}\ \,\sum_0^\infty \ \frac{x^{2n}}{(2n+4)!}\ [\,(2n+3)\,(2n+4)-\,8 +\,3\,(2n+4)\,]
\vspace{2mm}\\
&=&
\dis 
\,\sum_0^\infty \ \,\frac{6\,(n+1)\,(n+4)}{(2n+4)!}\ x^{2n} = 
\,1 + \frac{x^2}{12}+\frac{3\,x^4}{1\,120} +
 \frac{x^6}{21\,600} +\frac{x^8}{1\,995\,840} +\,...\ ,
 \vspace{2mm}\\
&\simeq &
\dis 
\,1 + .0833\ x^2 + 2.68\times 10^{-3}\ x^4 + 4.63 \times 10^{-5}\ x^6 + 5.01 \times 10^{-7}\ x^8 + \,...\ ,
\ea
\ee
providing back eq.\,(\ref{expphil}).
\vspace{2mm}

The corresponding effective density is
\be
\bar\rho_l(x) = \rho_0\ \frac{\Phi_l(x)}{\phi(x)}= \rho_0\ 
\frac{ (x^2\!-8)\cosh x +3\,x\sinh x+8}
{2x\,(x \,\cosh x -\sinh x)}\,,
\ee
decreasing regularly from $\bar\rho(0) = \rho_0$ down to $\rho(R)= \rho_0/2$ for higher $x$ corresponding to smaller $\lambda$'s.

\vspace{2mm}

While these expressions already provide good approximations of $\Phi(x)$ and $\bar\rho(x)$, as shown later in Table
\ref{tab:comp}, we can do better by averaging this linear density  $\rho_l(r)$ with the density $\rho_{-1}(r) =\rho_0 \ {2R}/{3r}$ considered earlier.

\vspace{-5mm}

\subsection{\boldmath A density $\,\rho'(r)=  \,\rho_0\dis \left(\,\frac54  - \frac{r}{R}+ \frac{R}{3r}\,\right)$}
\label{sec:dens}

\vspace{-2mm}

The linear expression (\ref{rhol0}) of the density distribution is simple and convenient, and gives a good approximation of $\Phi(x)$, as found in a 5-shell model, to within $1\,\%$ up to $x=5$. Still with $\rho(R) = \rho_0/2\simeq 2.75 $ g/cm$^3$ one has 
$\rho(.98 \,R)$ $ = .54\, \rho_0/2\simeq 2.98 $ g/cm$^3$  at a depth $R/50 \simeq 127 $ km, a density somewhat small compared to the outer mantle density ($\rho_{\rm om}\approx 3.5$ g/cm$^3$). This linearly decreasing density $\rho_l$ thus cannot pretend to represent very reliably $\Phi(x)$ for $x\approx 50$ (or  $\lambda \approx 130$ km), actually providing somewhat too small values of $\Phi(x)$ in this region, as shown in Table \ref{tab:comp}.

\begin{table}[t]
\caption{\ \,$\Phi(x)$ for several density distributions:
\vspace{-.4mm}
\,{\bf 1)} an homogeneous $\rho_0$, with \hbox{$\phi(x) = \frac{3}{x^3}\,(x\cosh x-\sinh x)$}; 
\hbox{{\bf 2)} $\rho_{-1}(r)= \rho_0\,\frac{2R}{3r}$}, 
\,with $\Phi_{-1}(x) = (\sinh \frac{x}{2} /\, \frac{x}{2})^2$;
\vspace{.2mm}
\,{\bf 3)} $\rho_l(r)= \rho_0\,(\frac52-2\,\frac{r}{R})$, 
\vspace{-.5mm}
with $\Phi_l(x) = \frac{3}{2x^4}\!
\,[(x^2\!-8)\cosh x + 3\, x\,\sinh x+8\,]$; 
\,{\bf 4)} $ \rho'(r)=\left(\frac54 -\frac{r}{R}+\frac{R}{3r}\right)$, 
with $\Phi'(x)= \frac{1}{4x^4}\,( \,7x^2\,\cosh x -24\,\cosh x + 9\,x\sinh x -4\,x^2+24\,)$;  \,{\bf 5)} in a 5-shell Earth model.
All densities 2) to 5) provide almost the same $\Phi(x)$ up to $x\simeq \, 4 $, 
to within $\approx 1\,\%$.
\,$\rho'(r)$ leads almost exactly to the same $\Phi(x)$ as the 5-shell model to within $\simeq .5\,\%$ up to $x=50$; and within $.7\,\%$ up to $x=64$ (i.e. 
$\lambda > 100$ km or $m< 2\times 10^{-12}$ eV$/c^2$). 
\vspace{-.1mm}
\,The effective density $\bar\rho\,'(x)$  is practically the same as in the 5-shell model (last column), both represented by the two superposed curves in blue, in Fig.\,\ref{rhorho}.
\label{tab:comp}
}
\vspace{4mm}
\label{tab:comp}
$\hspace{-6mm}\ba{|c||c|c|c||c|c||c|}
\hline
&&&&&&\vspace{-2mm}\\
\  \rho(r)\ & \rho_0&\ \rho_{-1} = \rho_0\ \hbox{\footnotesize$ \dis \frac{2R}{3r}$}\ & \, \rho_l = \rho_0 \left(\,\hbox{\footnotesize$ \dis \frac52$} -\!\hbox{\footnotesize$ \dis \frac{2r}{R}$}\,\right)  & 
 \, \rho'\! = \rho_0 \left(\hbox{\footnotesize$ \dis \frac54$} -\!\hbox{\footnotesize$ \dis \frac{r}{R}+\!\frac{R}{3r}$}\right) 
&\hbox{\small 5-shell model}& \ \dis  \frac{\bar\rho}{\rho_0}= \frac{\Phi}{\phi}\ 
\\
&&&&&&\vspace{-2mm}\\ 
\hline \hline
&&&&&&\vspace{-3mm}\\ 
\!\rho(R/2)\! &  \rho_0 &\frac43\,\rho_0& \frac32\,\rho_0& \frac{17}{12}\,\rho_0& 
\!\ba{l}
$\hbox{\footnotesize\ \ \  (too close to }$
\vspace{-1mm}\\
$\hbox{\footnotesize core/mantle disc.)}\!\!$
\ea\!
&\times\!\!\times\!\!\times\!\times
 \vspace{-3mm}\\ 
&&&&&&\vspace{0mm}\\ 
\hline
&&&&&&\vspace{-3mm}\\ 
\rho(R) &\rho_0& \frac23\,\rho_0 &  \frac12\,\rho_0  & \frac{7}{12}\,\rho_0& \frac12\,\rho_0 & \times\!\!\times\!\!\times\!\times \vspace{-3mm}\\ 
&&&&&&\vspace{0mm}\\ 
\hline\hline
&&&&&&
\vspace{-3mm}\\ 
\  \, x=1\ \,  &1.104&1.086 &1.086&1.086&1.085& .984\\
2&1.462&1.381&1.379& 1.380&1.377&.942\\
3&2.243&2.015&2.004& 2.010&2.002&.893\\
4&3.841&3.29&3.25& 3.27&3.25&.846\\
5&7.12&5.86&5.72&5.79&5.76&.808 \\
6&14.0&11.2&10.7 &10.9&10.9&.778 \\
8&61.1&46.5&43.7&45.1&45.0&.735
\vspace{-3.5mm}\\ &&&&&& \\
\hline
 &&&&&& \vspace{-3.5mm} \\
10&297&220&202& 211&211&.708 \\
20&1.73 \times 10^6\ &1.21 \times 10^6\ &1.03 \times 10^6\ &\ 1.12 \times 10^6\ \ &\  1.12 \times 10^6\ \ & .648\\
50&3.05\times 10^{18}&\,2.07\times 10^{18}\,&1.64\times 10^{18}&1.86\times 10^{18}&1.85 \times 10^{18} &.607\\
100&\ 3.99 \times 10^{39}\ &2.69 \times 10^{39}&2.07 \times 10^{39}&2.38 \times 10^{39} &2.33 \times 10^{39}&.584
\vspace{-3mm}\\
 &&&&&& \\
 \hline
\ea 
\hspace{-6mm}$
\vspace{2mm}
\end{table}

\begin{table}
\caption{ 
\,The ratios $r'$ and $r_{-1}$ of  $\,\Phi'(x)= [\,7x^2 \cosh x-24 \cosh x+ 9\,x\sinh x -4x^2+24\,]/(4\,x^4)$
and $\Phi_{-1}(x) = $ $(\sinh \frac{x}{2}/\frac{x}{2})^2\,$ (for the densities 
$\rho'\!=\rho_0\,(\frac54-\frac{r}{R}+\frac{R}{3r})$ and $ \rho_{-1}\!=\rho_0\,2R/3r$), relatively  to $\Phi_{\hbox{\footnotesize 5s}}(x)$  in a 5-shell model. $\Phi'(x)$ is almost equal to $\Phi_{\hbox{\footnotesize 5s}}(x)$, to within at most $.7\,\%$ for $x$ up to 64, i.e. $\lambda$ down to 100 km.
\label{tab:r}
\vspace{3mm}}
$
\ba{|c||c|c|c|c|c|c|c|c|c|c|c|c|c|c|}
\hline
&&&&&&&&&&&&&&\vspace{-3mm}\\ 
x&1&2&3&4&5&6&8&10 &15&20&30&50&64&100 \vspace{-2mm}\\ 
&&&&&&&&&&&&&&\vspace{-1mm}\\ 
\hline
&&&&&&&&&&&&&& \vspace{-3mm}\\ 
\ r'=\dis \frac{\Phi'(x)}{\Phi_{\hbox{\footnotesize 5s}}(x)}\ &1.001 & 1.002 & 1.004 & 1.005 & 1.005 &1.005& 1.003 & 1.001&.999 & 1.000 &1.002&1.004 & 1.007& 1.021
 \vspace{-3mm}\\ 
&&&&&&&&&&&&&&\vspace{0mm}\\ 
\hline
&&&&&&&&&&&&&& \vspace{-3mm}\\ 
\ r_{-1}=\dis \frac{\Phi_{-1}(x)}{\Phi_{\hbox{\footnotesize 5s}}(x)}\ &1.001 & 1.003 & 1.006 & 1.012 & 1.017 &1.024& 1.035 & 1.046 &1.066 &1.082 &1.102& 1.120& 1.129 & 1.153
 \vspace{-3mm}\\ 
&&&&&&&&&&&&&&\vspace{0mm}\\ 
\hline
\ea 
\vspace{2mm}
$
\end{table}

\vspace{2mm}
To overcome this inconvenience and better approach the result of the 5-shell calculation we shall consider a density distribution as the average
\be
\rho'(r) = \frac12\, \left(\,\rho_l(r)+\rho_{-1}(r) \,\right) \,= \,\rho_0\,\left(\,\frac54  - \frac{r}{R}+ \frac{R}{3r}\,\right)\ .
\ee
It corresponds to a slightly larger $\rho(R)= \frac{7}{12}\,\rho_0 \simeq 3.21$ g/cm$^3$ as intended, compensated by a slightly lower $\rho(R/2)= \frac{17}{12}\, \rho_0 \simeq 7.81$ g/cm$^3$, still with the same $I=\frac13\,MR^2$. The larger $\rho(R)$ as compared to the $\rho_0/2$ of $\rho_l$ (initially intended as representing $\rho_{\rm crust}$) is now closer to the outer mantle density, with the potential of providing a better approximation of $\Phi(x)$ and $\bar\rho(x)$ for smaller $\lambda$'s  down to 100 km corresponding to $x < 64\,$ or $m< 2\times 10^{-12} $ eV/$c^2$. The associated $\Phi(x)$ reads
\be
\label{Phi'}
\ba{c}
\Phi'(x) \,=\,\frac12\,\left(\Phi_{-1}(x)+ \Phi_l(x)\right) =  \frac{1}{x^2}\,(\cosh x-1) + \frac{3}{4x^4} \left[(x^2\!-8)\cosh x +3\,x\sinh x+8\,\right]\, ,
\ea
\ee
\vbox{
\noindent 
i.e.
\vspace{-1mm}
\be
\label{phin}
\framebox [10.6cm]{\rule[-.35cm]{0cm}{.95cm} $ \dis
\Phi'(x)\,=\, \dis \frac{1}{4x^4}\, ( \,7x^2\,\cosh x -24\,\cosh x + 9\,x\sinh x -4\,x^2+24\,)\ .
$}
\ee
}

\pagebreak

\noindent
This $\Phi'(x)$ now appears as a very good approximation of the 5-shell model result, as seen above 
in Tables \ref{tab:comp} and \ref{tab:r}, with
$\,.999 <\,\Phi'(x)/\Phi_{\hbox{\footnotesize 5s}}(x)$ $< 1.007\,$ for $x<64\,$ i.e. $\lambda$ down to 100 km.

\vspace{2mm}

The corresponding effective density is
\be
\framebox [10.2cm]{\rule[-.4cm]{0cm}{1.05cm} $ \dis
\bar\rho'(x) \,=\, \rho_0\ \frac{7x^2\,\cosh x -24\,\cosh x + 9\,x\sinh x -4\,x^2+24}{12\,x\,(x \,\cosh x -\sinh x)}\ ,
$}
\ee
behaving at large $x$ according to
$\,\bar\rho'(x)\to \frac{7}{12}\ \rho_0 = \rho'(R)$.

\vspace{2mm}
$\Phi'(x)$  may be expanded according to
\be
\ba{ccl}
\Phi '(x)\! &=&\dis \sum_0^\infty\  \left[\,\frac{1}{(2n+2)!}\,+\, \frac{3\,(n+1)(n+4)}{(2n+4)!}\,\right] \ x^{2n} = \,\sum_0^\infty \ \frac{7n^2+29n+24}{(2n+4)!} \ x^{2n}
\vspace{2mm}\\
 &=&\dis  1 + \frac{1}{12}\ x^2  + \frac{11}{4\,032} \ x^4+ \frac{29}{604\,800}\ x^6 + \frac{1}{1\,900\,800}\ x^8\,+\,...
\vspace{4mm}\\
&\simeq& 1+.0833\,x^2 + 2.73 \times 10^{-3}\,x^4+ 4.79 \times 10^{-5}\,x^6+ 5.26 \times 10^{-7}\,x^8+ 3.95 \times 10^{-9}\ x^{10}+\, ...\,. \!\!
\ea\!\!\!\!\!\!
\ee

\noindent
This is, again, remarkably close to the expansion of $\Phi(x)$ in a 5-shell model \cite{yuk}
\be
\label{exp5}
\Phi_{\hbox{\footnotesize 5s}}(x) \simeq \,1 +  .0827 \,x^2 + 2.71 \times 10^{-3}\,  x^4 + 4.78\times 10^{-5}\ x^6 +  5.26 \times 10^{-7}\ x^8 + 3.95 \times 10^{-9}\ x^{10} +\,...\, ,
\ee
as derived from its explicit expression
\be
\Phi_{\hbox{\footnotesize 5s}}(x)\,\simeq  \, .5\ \phi(x) +.1333\ \phi(.9953\ x)+.1715\ \phi(.8948 \ x)
+ .1929\ \phi(.5462\ x) + .0023\  \phi(.1916\ x)\ .
\ee

\begin{figure}[t]
\caption{\ The effective density $\bar\rho(x)=\rho_0\, \Phi(x)/\phi(x)$ as a function of $x=R/\lambda$ :  1) for $ \bar\rho =\rho_0$, in the homogeneous case (dotted green); 2)  for $\rho(r) = \rho_0 \,2R/3r$
with $\bar\rho(x) = \rho_0\ [\,2x\,(\cosh x -1)\,]/[\,3\,(x\cosh x-\sinh x)\,]$, decreasing to $\frac23\,\rho_0$ at large $x$  (dotted red);
3) for $\rho'(r) = \rho_0\,(\frac54-\frac{r}{R}+\frac{R}{3r})$ with $\bar\rho\,'(x) =  [ \,7x^2\,\cosh x -24\,\cosh x + 9\,x\sinh x -4\,x^2+24\,]\,/ \,[\,12x\,(x\,\cosh x-\sinh x)\,]$, decreasing to $\frac{7}{12}\,\rho_0$ (in blue);
4) in a 5-shell model, also in blue.
The last two curves almost exactly coincide, to within .7 \% for $x < 64$ i.e. $ \lambda > 100$ km.
\vspace{0mm}
\vspace{-.1mm}
The $1/r$ density profile also provides a good approximation of $\Phi(x)$ and $\bar\rho(x)$ (in dotted red) up to $x\simeq 4\,$.
\label{rhorho}
\vspace*{7mm}}
\includegraphics[width=10cm,height=7.2cm]{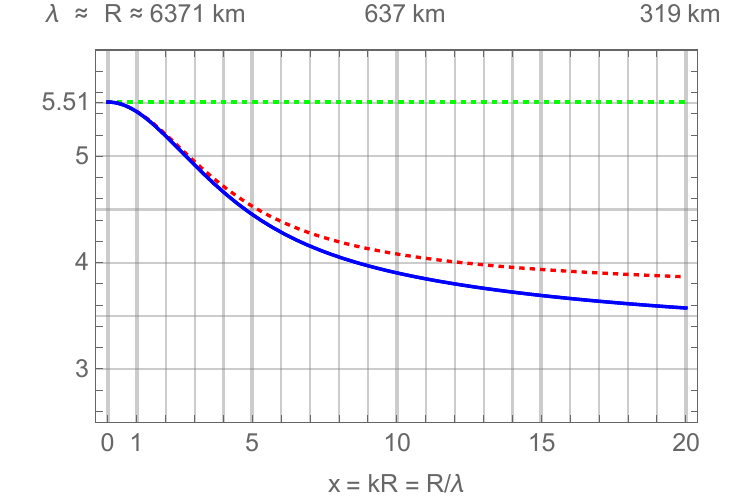}
\end{figure}

\vspace{2mm}

A comparison between these  hyperbolic form factors is given in Tables \ref{tab:comp} and \ref{tab:r}.
The distribution proportional to $1/r$
already provides a good approximation of $\Phi(x)$ and $\bar \rho(x)$, to within $1.2\,\%$ up to $x=4$.
The $\rho'(r)$ distribution provides analytic expressions $\Phi'(x)$ and $\bar \rho'(x)$ which almost coincide with the 5-shell model result, as seen in Fig.\,\ref{rhorho}, in which the curves corresponding to $\bar\rho\,'(x)$ and $\bar\rho_{\rm \,5s}(x)$, both in blue, are superposed.

\subsection{\boldmath  Limits on a new interaction effectively coupled to $B, \ L$, or $B-L$}

\begin{figure}[t]
\caption{\ Upper limits on $ |g_{B-L}|$ or $|g_L|$ (blue), and $ |g_B |$ (orange), at the 95\,\% CL. 
The limits for an E\" otv\" os parameter \hbox{$\delta  < 0$} are larger than for $\delta > 0$ by $\simeq 1.2$.
For $\lambda\gg R$ they are $1.1 \times  10^{-25}$ for $ |g_{B-L}|$ (spin-1) and $|g_L|$ (spin-0), solid blue line ; and
$1.3 \times  10^{-25}$ for $|g_L|$  (spin-1) and $ |g_{B-L}|$  (spin-0), dashed blue line. For $|g_B |$ they are $7.7\times 10^{-25}$ (spin 1, solid orange) and $6.4\times 10^{-25}$ (spin-0, dashed orange). 
For $m= 10^{-13},\, 10^{-12}$ or $10^{-11}$ eV/$c^2$ the limits are multiplied by $\simeq 1.9, \ 34$ or $1.2\times 10^9$, respectively, as compared to the nearly massless case.  
The limits obtained from $\Phi(x)$ in the 5-shell model or from the analytic expression $\Phi'(x)$ 
in eq.\,(\ref{phin}) differ by less than .4\,\% for $m < 2\times 10^{-12}$ eV/$c^2$, the corresponding curves being superposed.
\label{limitesg}
}
\vspace{3mm}
\hspace{4mm}\includegraphics[width=12cm,height=8.2cm]{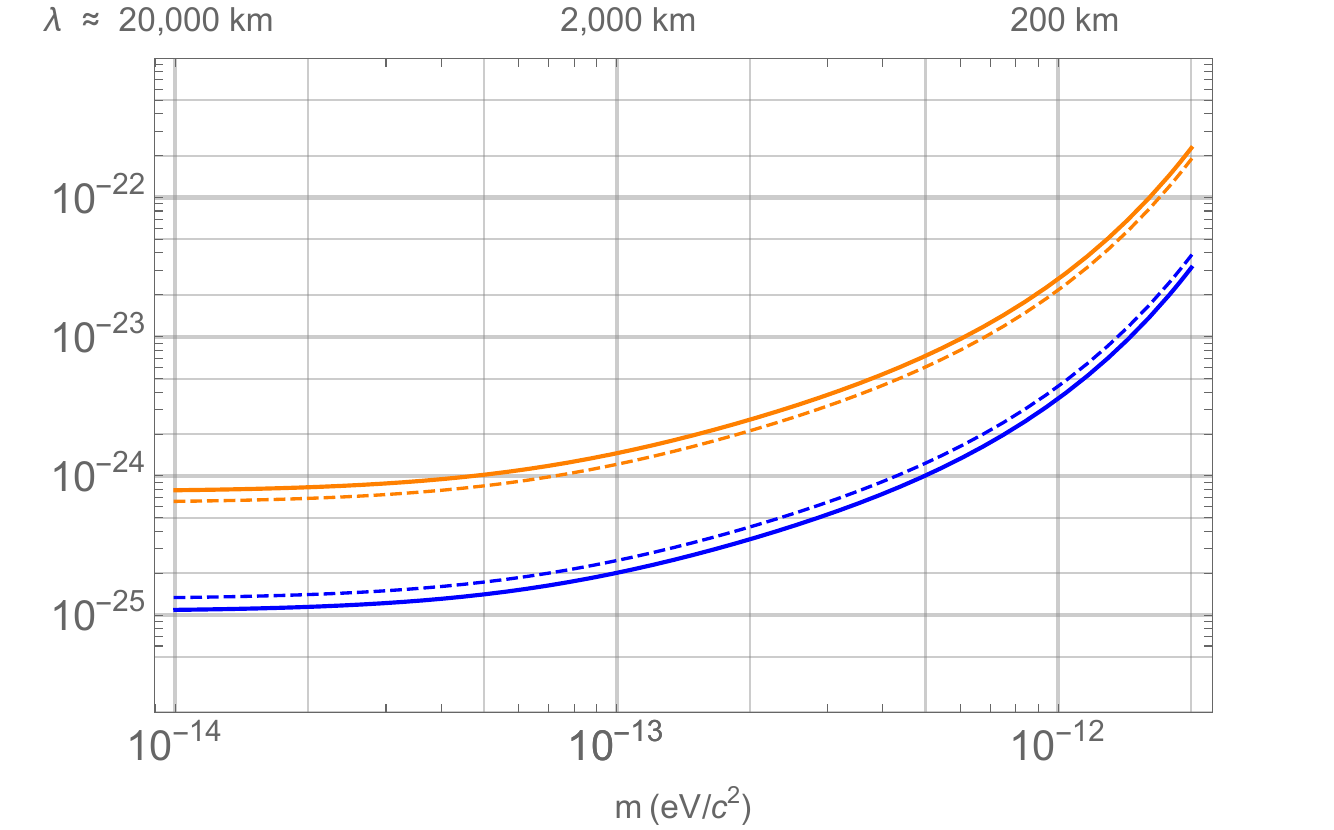}
\end{figure}

The ratio of a new force induced at a distance $r$ from the center by a mediator of mass $m=k$, as compared to a massless one, is given by
\be
F(x) \,=\, (1+kr) \,e^{-kr}\ \Phi(x) \,=\,  (1+kr) \ e^{-kr}\ \,\frac{3\,(x\,\cosh x-\sinh x)}{x^3}\ \,\frac{\bar\rho(x)}{\rho_0}\ .
\ee
The coupling limits obtained in the ultralight case i.e. for a range $\lambda$ significantly larger than the Earth radius
get rescaled by the following factor, inversely proportional to $\sqrt{\Phi(x)}$,
\be
F(x)^{-\frac12}\simeq 1.89,\  33.8, \  \hbox{or}\  1.24\times 10^9\,,\ \ \hbox{for} \ \ m \!=\! k\!= 10^{-13},\ 10^{-12},\ \hbox{or}\  10^{-11}\ \hbox{eV}/c^2\ .
\ee
$\Phi(x)$ in the 5-shell model is almost exactly given by the analytic expression 
$\Phi'(x)$ in eq.\,(\ref{phin}), to within .7\,\% up to $x=64$  corresponding to $\lambda >  100$ km, 
\vspace{-.4mm}
or to a mediator mass $m<  2\times 10^{-12}$ eV/$c^2$.
The coupling limits increase
with $m$ proportionally to $e^{mr/2}\!/\sqrt{\Phi(x) \,(1 + mr)}$, 
\vspace{-.4mm}
behaving at large $x$ like $mR\ e^{mz/2}/ \sqrt{1+mr}$.
\,They are shown in Fig.\,\ref{limitesg} for masses up to $2 \times 10^{-12}$ eV/$c^2$. In particular we have, for
$m=10^{-12}$ eV/$c^2$,
\be
\ba{c}
 \left\{\ 
\ba{rcl}
|g_{B-L}|_{\hbox{\scriptsize spin.-1}}\ \ [\,\hbox{or}\, \simeq |g_L|_{\hbox{\scriptsize spin-0}}\,]  <  \,3.6\times 10^{-24}\,,&& |g_{L}|_{\hbox{\scriptsize spin-1}}\  \ [\,\hbox{or}\, \simeq |g_{B-L}|_{\hbox{\scriptsize spin-0}} \,] <  \,4.4 \times 10^{-24}\ ,
\vspace{2mm}\\
|g_{B}|_{\hbox{\scriptsize spin-1}}  < \, 2.6\times 10^{-23}\,, && |g_{B}|_{\hbox{\scriptsize spin-0}}  <  \,2.2 \times 10^{-23}\ .
\ea\right.
\vspace{-3mm}\\
\ea
\ee

\section{Conclusion}

We discussed the general expressions of the inside and outside potentials generated by a sphere for a finite-range interaction, also taking advantage of extending the range of the radial coordinates, $r$ and $k$, to negative values. 
\vspace{.2mm}
The outside Yukawa potential is proportional to an hyperbolic form factor representative 
of the density distribution  $\rho(r)$. It  is 
\vspace{-.6mm}
given by the Laplace transform of the density distribution $\rho(\vec r)$, $\,\Phi(x)= \langle\, \cosh\,\vec k.\vec r\,\rangle= \langle\,\frac{\sinh kr}{kr}\,\rangle$,
\vspace{-.9mm}
 related by duality or analytic continuation to the ordinary form factor $\Phi(ix)= \langle\, e^{i\,\vec k.\vec r}\,\rangle= \langle\,\frac{\sin kr}{kr}\,\rangle$, with $x=kR=R/\lambda$.
\,It  is also obtained from the bilateral Laplace transform of $2\pi r \rho(|r|)$, allowing one to recover the value of the density $\rho(r)$ from an integral over $\Phi(ix)$. 
 \,We have evaluated $\Phi(x)$, and their power series expansions involving the even moments $\langle\, r^{2n}\,\rangle$, for various density distributions. We expressed them relatively to the $\phi(x)$ for an homogeneous sphere, in terms of effective densities $\bar\rho(x)= \rho_0\,\Phi(x)/\phi(x)$, decreasing, for $d\rho /dr \!<0\,$, from the average density $\rho_0$ at small $x$ down to the surface density $\rho(R)$ at large $x$.

\vspace{2mm}

The limits on an extremely weak new long range force induced by an ultralight mediator of mass $m=1/\lambda$, as deduced from the {\it MICROSCOPE\,} experiment,
 \vspace{-.5mm}
 increase rapidly with the mediator mass, and are inversely proportional to $\sqrt{\Phi(x)}$\,.
Quite remarkably, $\Phi(x)$ is not so sensitive to the details of the mass distribution within the interior of the Earth. 

\vspace{2mm}
We have concentrated on two simple density distributions, providing analytic expressions of their hyperbolic form factors $\Phi(x)$, and compared them with the one obtained in a 5-shell Earth model distinguishing between inner and outer cores, inner and outer mantles, and crust.
 \vspace{-.5mm}
The density $\rho(r) = \rho_0\,\frac{2R}{3r}$ leads to a very simple $\Phi(x)= \frac{2}{x^2}\,(\cosh x -1)$, 
 \vspace{-.1mm}
valid to within $1.2\,\%$ up to $x= 4$ (and 5\,\% up to $x=10$),
 \vspace{-.4mm}
with  $\Phi(x) =  1 + \frac{1}{12}\, x^2  + \frac{1}{360} \, x^4 $ $+\, ...\ $. 
\,The other is $\rho'(r) = \rho_0 \,(\frac54 - \frac{r}{R}+\frac{R}{3r})$,
\vspace{.2mm}
whose $\Phi'(x)$ reproduces remarkably well the result of the 5-shell model,
to within .7\,\% up to $x= 64$ or $\lambda$ down to 100 km,  with an expansion $\Phi'(x)= 1 + \frac{1}{12}\, x^2 + \frac{11}{4\,032}\, x^4+  \, ...\ $.

\vspace{2mm}

In practice though, the simpler expression
\vspace{-5mm}

\be
\sqrt{\Phi(x)}\  \simeq \ \frac{\sinh x/2}{x/2}\ ,
\ee
\vspace{-4mm}

\noindent
valid for the Earth to within $\simeq 2.3$ \% for $\lambda$ down to $R/10$\,, 
may be sufficient in most cases.
In addition, the general properties of the hyperbolic form factors discussed here apply to other situations involving finite-range interactions.

\bibliography{References}

\end{document}